\documentclass[aps,prd,reprint,superscriptaddress,nofootinbib,floatfix]{revtex4-2}
\usepackage{amsmath,amssymb,bm}
\usepackage{graphicx,booktabs,array}
\usepackage{xcolor}
\usepackage{tikz-feynman}
\tikzfeynmanset{warn luatex=false,compat=1.1.0}
\usepackage{overpic}
\usepackage[colorlinks=true,linkcolor=blue,citecolor=blue,urlcolor=blue]{hyperref}
\graphicspath{{figures/}}
\newcommand{\pt}{p_T}
\newcommand{\msd}{m_{\mathrm{SD}}}
\newcommand{\ip}{\mathrm{IP}}
\newcommand{\fb}{\mathrm{fb}^{-1}}
\newcommand{\GeV}{\mathrm{GeV}}
\newcommand{\jbb}{J_1^{2b}}
\newcommand{\pjet}{p_T^{\mathrm{jet}}}
\newcommand{\displaced}{0.1<\ip_{3D}<10~\mathrm{mm}}
\newcommand{\ndisplaced}{N_{\mathrm{trk}}(\displaced)}
\newcommand{\fdisplaced}{\frac{\sum p_T^{\mathrm{trk}}}{p_T^{\mathrm{jet}}}(\displaced)}
\begin{document}

\title{Squeezed double-\texorpdfstring{$b$}{b} tagging for light hadronic resonances at the LHC}
\author{Dilip Kumar Ghosh}
\email{tpdkg@iacs.res.in}
\affiliation{School of Physical Sciences, Indian Association for the Cultivation of Science, 2A \& 2B Raja S. C. Mallick Road, Jadavpur, Kolkata 700032, India}
\author{Biswarup Mukhopadhyaya}
\email{biswarup@iiserkol.ac.in}
\affiliation{School of Physical Sciences, Indian Association for the Cultivation of Science, 2A \& 2B Raja S. C. Mallick Road, Jadavpur, Kolkata 700032, India}
\affiliation{Department of Physical Sciences, Indian Institute of Science Education and Research Kolkata, Mohanpur 741246, India}
\author{Sirshendu Samanta}
\email{ss21rs027@iiserkol.ac.in}
\affiliation{Department of Physical Sciences, Indian Institute of Science Education and Research Kolkata, Mohanpur 741246, India}
\author{Ritesh K. Singh}
\email{ritesh.singh@iiserkol.ac.in}
\affiliation{Department of Physical Sciences, Indian Institute of Science Education and Research Kolkata, Mohanpur 741246, India}

\begin{abstract}
A boosted light resonance decaying into $b\bar b$ can produce a single jet containing two bottom-hadron decay systems. We investigate this squeezed double-$b$ topology using a tracking-based boosted decision tree that distinguishes non-bottom, single-bottom and double-bottom jets. The classifier combines displaced-track multiplicities and momentum fractions with ordered impact parameters and jet kinematics. For the simulated jet samples, the double-bottom tagging response is approximately $83\%$, with corresponding single-bottom and non-bottom mistag responses of $9.7\%$ and $0.5\%$; charm contamination is assessed separately. The tagger is combined with soft-drop mass reconstruction and an event-level classifier using the candidate and recoil jets to identify light hadronic resonances within the QCD background. As an illustrative example, we consider a singlet extension of the flipped two-Higgs-doublet model with pseudoscalar masses of 30, 50 and 60 GeV. The signal samples use a simplified effective-vertex approximation, replacing the momentum-dependent quark-loop production interaction by a local vertex. A resolved loop calculation and higher-order production corrections are beyond the scope of this tagging study. The resulting reconstruction strategy can be applied to different signal hypotheses with their corresponding kinematic acceptance and tagging response.
\end{abstract}
\maketitle

\section{Introduction}
\label{sec:introduction}

Hadronic decays of a light resonance are difficult to isolate at the LHC because of the large QCD multijet background. Decays into $b\bar b$ provide displaced tracks and heavy-flavour decay vertices, but conventional searches based on two separate bottom jets can lose efficiency when the resonance is boosted. The two decay systems may then lie inside one jet. Recovering their bottom-hadron multiplicity, rather than resolving two jet axes, becomes the relevant reconstruction problem.

Hard recoil radiation provides a useful way to access this regime. The accompanying radiation gives transverse momentum to the resonance, raising the momentum scale of the event while bringing the resonance's decay products closer together. For a sufficiently light state, the resulting compact decay can therefore be an essential part of the search signature. The same recoil that motivates the event selection also makes identifying two bottom-hadron decay systems within one jet a central requirement.

We call this configuration a squeezed double-$b$ jet. The term refers to a compact pair of bottom-hadron decay systems contained within a jet of moderate radius. It describes an observable topology and does not identify a particular production mechanism or scalar potential. Existing double-$b$ tagging studies already establish the usefulness of heavy-flavour information in boosted jets~\cite{CMS:2016doubleb}. Here we investigate a tracking-based three-category tagger for light resonances, retaining the distinction between non-bottom, single-bottom and double-bottom configurations throughout the analysis. The two-bottom category must be separated both from ordinary jets with no bottom hadron and from jets with one bottom hadron, whose displaced tracks can otherwise imitate part of the signal.

The background also contains genuine compact bottom-quark pairs from QCD gluon splitting. Such jets can belong to the same $2b$ category as the resonance decay, so identifying their bottom-hadron content is one stage of the search. The reconstructed mass and the surrounding event kinematics supply additional information needed to isolate a resonant contribution within this population.

The central observables are the multiplicity and transverse momentum carried by displaced tracks, together with several ordered impact parameters. A single large displacement is insufficient to distinguish a jet containing two bottom hadrons from a jet containing one. The pattern across multiple tracks instead supplies information about bottom-hadron multiplicity. We use boosted decision trees (BDTs)~\cite{Roe:2004na}, implemented with \texttt{XGBoost}~\cite{chen2016xgboost}, for this classification. Jet mass is subsequently reconstructed with soft drop~\cite{Larkoski:2014wba}, and an event-level BDT uses the tagged candidate and its recoil system to reduce the remaining background.

The study follows this sequence from jet classification to event selection. We examine the discriminating tracking variables, evaluate the bottom- and charm-jet responses, and compare the groomed-mass spectra before and after classification. The event analysis then combines the reconstructed candidate with its recoil system. Presenting these stages together makes their respective contributions explicit and connects the squeezed-jet reconstruction to a search for a light hadronic resonance.

Our target can be written generically as
\begin{equation}
 pp\to \phi+j(j),\qquad \phi\to b\bar b,
 \label{eq:generic-process}
\end{equation}
where $\phi$ denotes a light spin-zero resonance and $j(j)$ denotes one or two hard recoil partons. Neither the double-$b$ tag nor the jet-mass reconstruction requires knowledge of the other particles in the underlying theory. The acceptance and classification efficiencies nevertheless depend on the resonance mass, boost and radiation pattern. The reconstruction strategy can be applied to different signal hypotheses, with the acceptance and tagging efficiencies evaluated for the corresponding kinematics and the production rate specified separately.

For a concrete illustration we use a light pseudoscalar in a singlet-extended flipped two-Higgs-doublet model (2HDM)~\cite{Branco:2011iw,Bhattacharyya:2015nca,Arcadi:2020gge,Arcadi:2022lpp}. This supplies a convenient set of hadronic signal samples. The analysis does not use the heavy doublet states to identify the signal and cannot establish the flipped Yukawa assignment from the squeezed jet alone. The production sample is described by a simplified local effective vertex, whose role and limitations are stated explicitly below.

Low-mass boosted-resonance searches provide the experimental context. CMS has searched for boosted $b\bar b$ resonances between 50 and 350 GeV using $35.9~\fb$ at 13 TeV~\cite{CMS:2018pwl}. A subsequent CMS search uses $138~\fb$ to study quark-pair resonances between 50 and 300 GeV~\cite{CMS:2026lowmass}. ATLAS has also explored lower hadronic masses in a photon-associated channel~\cite{ATLAS:2025lowmass}. These analyses motivate attention to low masses and compact jets, while their different recoil, tagging and trigger requirements must be respected when comparing rates.

Section~\ref{sec:topology} introduces the squeezed topology and its rate interpretation. Section~\ref{sec:model} defines the illustrative model and benchmarks. Simulation and reconstruction are described in Sec.~\ref{sec:simulation}, followed by the tagger and its performance in Sec.~\ref{sec:tagger}. Sections~\ref{sec:event} and \ref{sec:sensitivity} present the event selection and conditional sensitivity. We conclude in Sec.~\ref{sec:conclusion}.

\section{The squeezed double-\texorpdfstring{$b$}{b} topology}
\label{sec:topology}

For two approximately massless decay products, their invariant mass satisfies
\begin{equation}
 m_{b\bar b}^2=2\pt^b\pt^{\bar b}
 \left[\cosh(\Delta\eta)-\cos(\Delta\phi)\right].
\end{equation}
In the small-angle, boosted limit, writing $z$ for the transverse momentum fraction of one daughter gives
\begin{equation}
 m_{b\bar b}^2\simeq z(1-z)(\pt^{b\bar b})^2\Delta R_{b\bar b}^{\,2},
 \qquad \Delta R=\sqrt{\Delta\eta^2+\Delta\phi^2}.
\end{equation}
For a roughly symmetric decay this becomes
\begin{equation}
 \Delta R_{b\bar b}\simeq\frac{2m_\phi}{\pt^\phi}.
 \label{eq:opening}
\end{equation}
At $\pt^\phi=200$ GeV, masses of 30, 50 and 60 GeV correspond to characteristic separations of 0.3, 0.5 and 0.6, respectively. These values show both the opportunity and the acceptance limitation of a jet with radius $R=0.5$: the lightest state is readily squeezed, whereas the heavier states generally need a larger boost for both decay systems to be contained. The relation is approximate; momentum sharing, showering and reconstruction broaden it.

Recoil radiation is useful because it supplies the transverse momentum needed to enter this regime. A hard recoil can also support experimental triggering, although a detector-level trigger efficiency is not inferred from an offline jet requirement alone. Within the candidate jet, two bottom hadrons yield multiple displaced decay tracks. The tagger uses that structure, while the groomed jet mass supplies an estimate of the parent mass. These are distinct tasks: classification into the $0b$, $1b$ and $2b$ jet categories need not by itself produce a narrow resonance peak, and a mass peak does not measure a double-$b$ tagging efficiency.

For a new signal hypothesis the selected yield can be expressed as
\begin{equation}
 S=\mathcal L\,\sigma(pp\to\phi+X)\,
 \mathrm{BR}(\phi\to b\bar b)\,A\,
 \epsilon_{\mathrm{reco}}\epsilon_{\mathrm{tag}}\epsilon_{\mathrm{sel}}.
 \label{eq:yield-factorization}
\end{equation}
Here $S$ is the expected number of selected signal events and $\mathcal L$ is the integrated luminosity. The production cross section $\sigma(pp\to\phi+X)$ describes production of the light resonance $\phi$ with accompanying radiation or other particles denoted by $X$. The branching fraction $\mathrm{BR}(\phi\to b\bar b)$ is the probability that $\phi$ decays into a bottom quark and antiquark. The factor $A$ is the geometric and kinematic acceptance of the signal. Within this acceptance, $\epsilon_{\mathrm{reco}}$ is the efficiency for reconstructing the required objects, $\epsilon_{\mathrm{tag}}$ is the efficiency for identifying the candidate as a double-$b$ jet, and $\epsilon_{\mathrm{sel}}$ is the efficiency of the remaining event selection, including the mass window and event-BDT requirement. Each efficiency is conditional on the preceding stages, so their product with $A$ represents the total selection efficiency without assuming that the stages are independent.

The acceptance and efficiencies can depend on $m_\phi$, $\pt^\phi$, additional radiation and detector conditions. The visible cross section is $\sigma_{\mathrm{vis}}=S/\mathcal L$. This formulation separates the reconstructed squeezed object from its assumed production rate. Rescaling a benchmark rate is meaningful at fixed signal kinematics and response; transferring the result to a different spectrum requires the corresponding acceptance and efficiency evaluation.

\section{An illustrative light-pseudoscalar model}
\label{sec:model}

\subsection{Scalar sector and Yukawa interactions}

We take a CP-conserving flipped 2HDM and add a real pseudoscalar singlet $P$. This construction provides two CP-odd mass eigenstates, $A$ and $a$, with the lighter state used in the collider illustration. The scalar potential is
\begin{equation}
 V=V_{\mathrm{2HDM}}+V_P,
\end{equation}
with
\begin{align}
 V_{\mathrm{2HDM}}={}&m_{11}^2\Phi_1^\dagger\Phi_1
 +m_{22}^2\Phi_2^\dagger\Phi_2
 -[m_{12}^2\Phi_1^\dagger\Phi_2+\mathrm{h.c.}]
 \nonumber\\
 &+\frac{\lambda_1}{2}(\Phi_1^\dagger\Phi_1)^2
 +\frac{\lambda_2}{2}(\Phi_2^\dagger\Phi_2)^2
 \nonumber\\
 &+\lambda_3(\Phi_1^\dagger\Phi_1)(\Phi_2^\dagger\Phi_2)
 +\lambda_4(\Phi_1^\dagger\Phi_2)(\Phi_2^\dagger\Phi_1)
 \nonumber\\
 &+\left[\frac{\lambda_5}{2}(\Phi_1^\dagger\Phi_2)^2+\mathrm{h.c.}\right],
 \label{eq:potential-doublet}
\end{align}
and
\begin{align}
 V_P={}&\frac12m_P^2P^2+\frac{\lambda_P}{4}P^4
 +P^2[\kappa_1\Phi_1^\dagger\Phi_1+\kappa_2\Phi_2^\dagger\Phi_2]
 \nonumber\\
 &+i\kappa_3P(\Phi_1^\dagger\Phi_2-\Phi_2^\dagger\Phi_1).
 \label{eq:potential-singlet}
\end{align}
The parameters are real, $v=\sqrt{v_1^2+v_2^2}$ is the electroweak vacuum expectation value, and $\tan\beta=v_2/v_1$. In the $(A_{\mathrm{2HDM}},P)$ basis the CP-odd squared-mass matrix is
\begin{equation}
 \mathcal M_P^2=\begin{pmatrix}m_{AA}^2&m_{AP}^2\\m_{AP}^2&m_{PP}^2\end{pmatrix},
\end{equation}
where
\begin{align}
 m_{AA}^2&=\frac{m_{12}^2}{\sin\beta\cos\beta}-v^2\lambda_5,\nonumber\\
 m_{PP}^2&=m_P^2+(\kappa_1\cos^2\beta+\kappa_2\sin^2\beta)v^2,\nonumber\\
 m_{AP}^2&=-\kappa_3v.
\end{align}
The physical states are defined by
\begin{equation}
 \begin{pmatrix}A\\a\end{pmatrix}=
 \begin{pmatrix}\cos\theta&-\sin\theta\\\sin\theta&\cos\theta\end{pmatrix}
 \begin{pmatrix}A_{\mathrm{2HDM}}\\P\end{pmatrix},
 \label{eq:rotation}
\end{equation}
with
\begin{align}
 m_{A,a}^2={}&\frac12\left[m_{AA}^2+m_{PP}^2
 \vphantom{\sqrt{(m_{AA}^2)^2}}\right.
 \nonumber\\[-1mm]
 &\left.\hspace{6mm}\pm\sqrt{(m_{AA}^2-m_{PP}^2)^2+4(m_{AP}^2)^2}\right],
 \label{eq:eigenvalues}\\
 \tan2\theta={}&\frac{-2m_{AP}^2}{m_{AA}^2-m_{PP}^2}.
\end{align}

The doublet sector also contains two CP-even states, $h$ and $H$, and a charged pair $H^\pm$. We identify $h$ with the observed Higgs boson. In the absence of the singlet, the charged--CP-odd doublet splitting obeys
\begin{equation}
 m_{H^\pm}^2-m_{AA}^2=\frac{v^2}{2}(\lambda_5-\lambda_4).
 \label{eq:doublet-splitting}
\end{equation}
A very light doublet pseudoscalar together with a heavy charged scalar can therefore require sizeable quartic couplings. The singlet extension separates the physical light eigenvalue from $m_{AA}^2$, relaxing this tension. It does not remove the need to check stability and perturbative unitarity for each parameter choice.

In the flipped Yukawa assignment, $\Phi_2$ couples to up-type quarks and charged leptons, while $\Phi_1$ couples to down-type quarks. For the light state,
\begin{equation}
 \mathcal L_Y^a=-i\sum_f\frac{m_f}{v}\xi_f^a\bar f\gamma_5f\,a,
 \label{eq:yukawa}
\end{equation}
where
\begin{equation}
 \xi_u^a=\sin\theta\cot\beta,\quad
 \xi_d^a=\sin\theta\tan\beta,\quad
 \xi_\ell^a=-\sin\theta\cot\beta.
\end{equation}
The mixing also reduces the doublet-mediated electroweak production couplings of $a$. Associated $Za$ production, studied for a light doublet pseudoscalar in Ref.~\cite{Ghosh:2025kju}, can consequently be suppressed. Hadronic recoil production provides a different way to obtain a boosted $b\bar b$ system, although it retains the coupling dependence of the underlying production interaction. The quark couplings scale differently with $\tan\beta$; no universal production-rate enhancement follows from the topology alone.

\subsection{Constraints and benchmark choice}
\label{sec:constraints}

The illustrative parameter points are selected using the theoretical and experimental tests described below. The doublet boundedness conditions include~\cite{Nie:1998yn,Arcadi:2020gge}
\begin{align}
 &\lambda_1>0,\qquad\lambda_2>0,\qquad
 \lambda_3>-\sqrt{\lambda_1\lambda_2},\nonumber\\
 &\lambda_3+\lambda_4-|\lambda_5|>-\sqrt{\lambda_1\lambda_2}.
\end{align}
Necessary conditions involving the singlet directions are
\begin{equation}
 \lambda_P>0,\qquad
 \kappa_i>-\sqrt{\frac{\lambda_i\lambda_P}{2}}\quad(i=1,2).
\end{equation}
These pairwise inequalities alone are not a sufficient boundedness test when several fields are simultaneously nonzero; the mixed directions must also be checked. Tree-level scalar scattering is required to satisfy $|\Lambda_i|<8\pi$, including the singlet contributions to the scattering matrices~\cite{PhysRevD.7.3111,PhysRevD.16.1519}. The relevant electroweak vacuum is checked against competing minima. These are consistency requirements at the scales of the analysis, rather than a statement about validity up to an arbitrarily high scale.

The collider checks use \texttt{HiggsBounds} and \texttt{HiggsSignals}~\cite{Bechtle:2020pkv,Bechtle:2020uwn,Bahl:2022igd}. They include searches for additional scalars and exotic Higgs decays, as well as agreement of the 125 GeV signal strengths within $2\sigma$. Flavour tests include the radiative decay $b\to s\gamma$, using the reference branching fraction $(3.32\pm0.15)\times10^{-4}$~\cite{HFLAV:2016hnz}, and $B_s\to\mu^+\mu^-$~\cite{CMS:2014xfa}. Charged-Higgs limits depend on the Yukawa structure and $\tan\beta$~\cite{Misiak:2017flavour}; the selected low-$\tan\beta$ examples have charged masses above approximately 600 GeV. We also require compatibility with the reference 95\% confidence region for the oblique parameters $S$, $T$ and $U$~\cite{PhysRevD.46.381,ALEPH:2005ab,10.1093/ptep/ptaa104}. These checks define the parameter selection used for the illustration.

\begin{figure*}[t]
 \centering
 \includegraphics[width=0.485\textwidth]{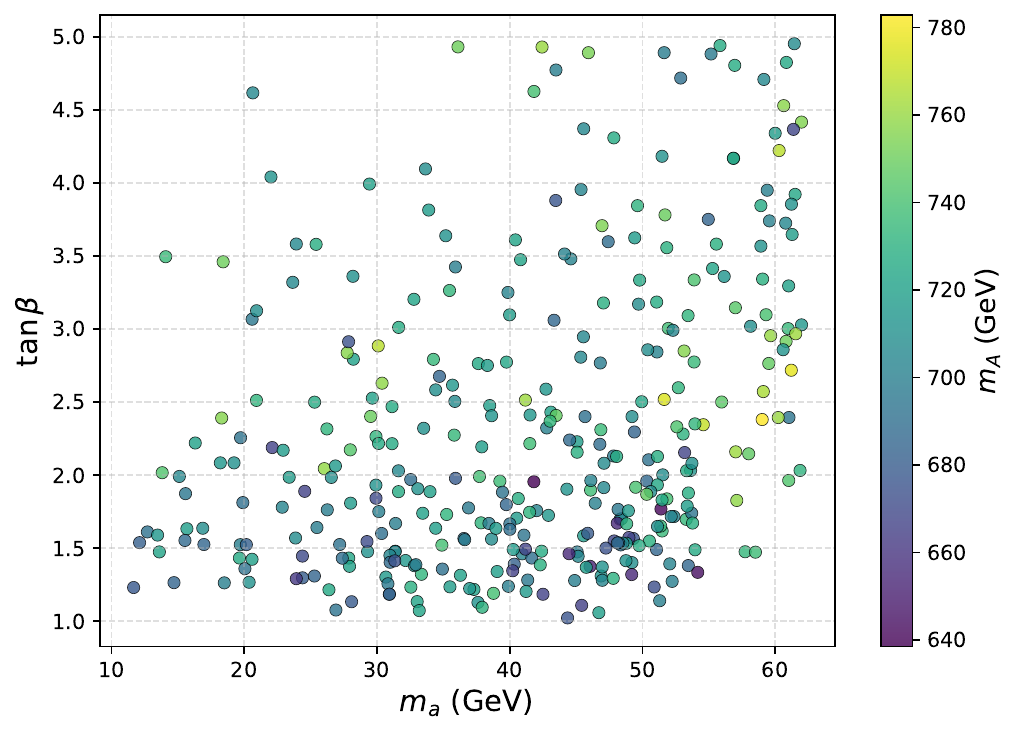}\hfill
 \includegraphics[width=0.485\textwidth]{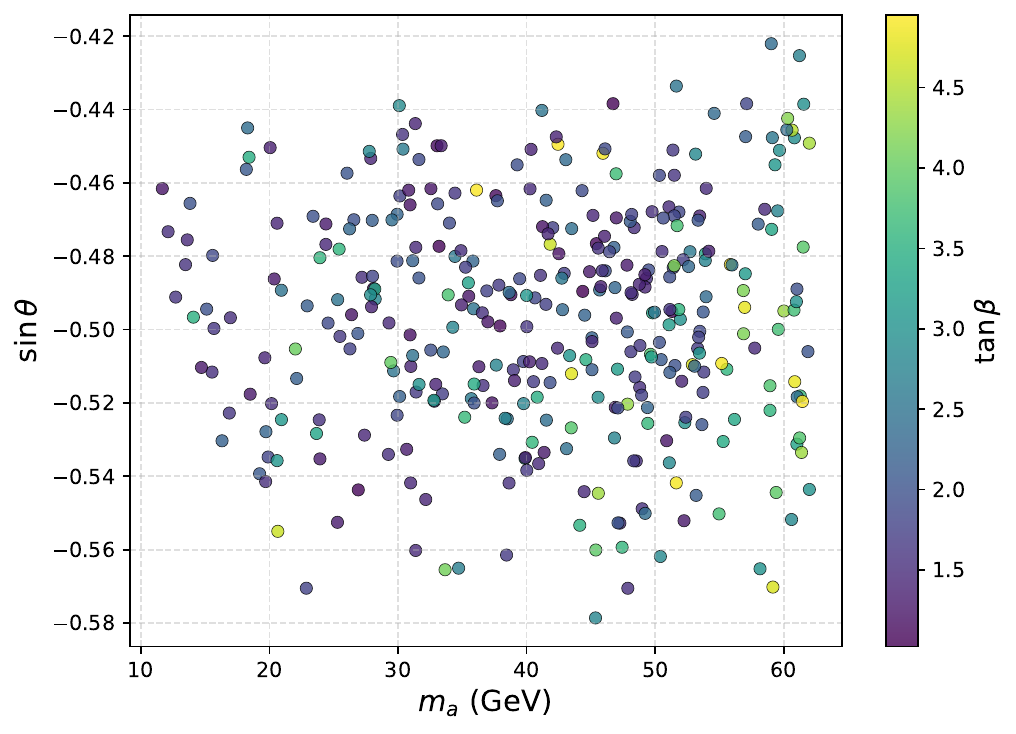}
 \caption{Illustrative parameter points passing the model-selection tests. Left: projection onto $m_a$ and $\tan\beta$, with the heavy CP-odd mass $m_A$ shown by the colour scale. Right: the corresponding $m_a$--$\sin\theta$ projection.}
 \label{fig:constraints_scan}
\end{figure*}

Figure~\ref{fig:constraints_scan} retains the parameter-space projections used to select the examples. A populated band in such a projection should not be read as a measurement of $\tan\beta$ or as an exclusion of all unpopulated points. In particular, the sign of $\theta$ depends on the singlet-field convention. Under $P\to-P$, accompanied by $\kappa_3\to-\kappa_3$, one may reverse $\theta$ and the sign of the light eigenstate without changing the rates considered here. The negative branch shown in the figure is therefore not a sign preference established by flavour or collider data. The limit $\sin\theta\to0$ is a decoupling limit for the light singlet's visible couplings, not a region excluded merely by its absence from the plotted points.

\begin{table}[t]
 \caption{Benchmark points of the illustrative singlet-extended flipped two-Higgs-doublet model, selected to study squeezed double-$b$ reconstruction at different pseudoscalar masses. All masses are given in GeV.}
 \label{tab:benchmarks}
 \centering
 \begin{ruledtabular}
 \begin{tabular}{lrrrrr}
 & $m_a$ & $m_A$ & $m_{H^\pm}$ & $\tan\beta$ & $\sin\theta$\\
 BP1 & 30 & 703 & 609 & 1.6 & $-0.58$\\
 BP2 & 50 & 705 & 608 & 1.7 & $-0.57$\\
 BP3 & 60 & 675 & 647 & 1.4 & $-0.45$
 \end{tabular}
 \end{ruledtabular}
\end{table}

The three benchmarks in Table~\ref{tab:benchmarks} have $m_a=30$, 50 and 60 GeV. Our main purpose in choosing them is to probe how the pseudoscalar mass changes the opening angle, jet containment and mass reconstruction. The nearby values of the other parameters keep the examples within a similar coupling regime. They are not intended to span all allowed couplings or demonstrate a reach across the entire scalar-sector parameter space. In particular, a larger-$\tan\beta$ signal requires its own production normalization and kinematic evaluation.

\section{Signal simulation and jet reconstruction}
\label{sec:simulation}

\subsection{Simplified effective-vertex production}

The illustrative signal is
\begin{equation}
 pp\to a+j(j),\qquad a\to b\bar b.
 \label{eq:signal-process}
\end{equation}
The events use a \texttt{SARAH}-generated model implementation~\cite{Staub:2013tta} with a local gluon--spin-zero effective interaction. This is a simplified production description used to generate the squeezed topology. It replaces a momentum-dependent quark-loop production subamplitude by a local vertex. The sample therefore does not constitute an evaluation of the resolved loop amplitudes for a light pseudoscalar recoiling against hard partons.

\begin{figure*}[t]
    \centering
    \resizebox{0.48\textwidth}{!}{
        \begin{tikzpicture}
            \begin{feynman}
                \vertex (q1) {\(q\)};
                \vertex [below=3cm of q1] (g2) {\(g\)};
                \vertex [right=1.5cm of q1] (isr_v);
                \vertex [above right=1.2cm of isr_v] (split_q);
                \vertex [above=0.8cm of split_q] (out_q) {\(q\)};
                \vertex [right=1.0cm of split_q] (out_g) {\((g)\)};
                \vertex [right=2.0cm of isr_v, yshift=-1.5cm] (t1); 
                \vertex [below=1.5cm of t1] (t2); 
                \vertex [right=1.5cm of t1, yshift=-0.75cm] (t3);
                
                \vertex [right=1.5cm of t3] (a1) {\(a\)};
                \vertex [right=1.5cm of a1, yshift=0.3cm] (b) {\(b\)};
                \vertex [right=1.5cm of a1, yshift=-0.3cm] (bbar) {\(\bar{b}\)};

                \diagram* {
                    (q1) -- [fermion] (isr_v),
                    (g2) -- [gluon] (t2),    
                    
                    (isr_v) -- [fermion] (split_q),
                    (split_q) -- [fermion] (out_q),
                    (split_q) -- [gluon, thick] (out_g),
                    
                    (isr_v) -- [gluon] (t1),
                    (t1) -- [fermion] (t3),
                    (t3) -- [fermion] (t2),
                    (t2) -- [fermion] (t1),
                    (t3) -- [scalar] (a1),
                    (a1) -- [fermion] (b),
                    (a1) -- [anti fermion] (bbar),
                };
            \end{feynman}
        \end{tikzpicture}
    }
    \hfill
    \resizebox{0.48\textwidth}{!}{
        \begin{tikzpicture}
            \begin{feynman}
                \vertex (g1) {\(g\)};
                \vertex [below=3cm of g1] (g2) {\(g\)};
                \vertex [right=1.5cm of g1] (isr_v);
                \vertex [above right=1.2cm of isr_v] (split_v);
                \vertex [above=0.8cm of split_v] (j1) {\((g)\)};
                \vertex [right=1.0cm of split_v] (j2) {\(g\)};
                
                \vertex [right=2.0cm of isr_v, yshift=-1.5cm] (t1);
                \vertex [below=1.5cm of t1] (t2);
                \vertex [right=1.5cm of t1, yshift=-0.75cm] (t3);
                
                \vertex [right=1.5cm of t3] (a1) {\(a\)};
                \vertex [right=1.5cm of a1, yshift=0.3cm] (b) {\(b\)};
                \vertex [right=1.5cm of a1, yshift=-0.3cm] (bbar) {\(\bar{b}\)};

                \diagram* {
                    (g1) -- [gluon] (isr_v),
                    (isr_v) -- [gluon] (t1), 
                    (g2) -- [gluon] (t2),    
                    
                    (isr_v) -- [gluon] (split_v),
                    (split_v) -- [gluon, thick] (j1),
                    (split_v) -- [gluon] (j2),
                    
                    (t1) -- [fermion] (t3),
                    (t3) -- [fermion] (t2),
                    (t2) -- [fermion] (t1),
                    
                    (t3) -- [scalar] (a1),
                    
                    (a1) -- [fermion] (b),
                    (a1) -- [anti fermion] (bbar),
                };
            \end{feynman}
        \end{tikzpicture}
    }
    \caption{Representative triangle-induced configurations for $pp\to a+j(j)$, followed by $a\to b\bar b$, showing quark-initiated (left) and gluon-initiated (right) recoil production. Parentheses indicate an additional radiated gluon. The quark-loop production interaction is treated using the simplified effective-vertex approximation in the signal simulation.}
    \label{fig:feynman_diag}
\end{figure*}

Figure~\ref{fig:feynman_diag} shows representative triangle-induced configurations motivating this description. It illustrates how radiation supplies recoil and how the light state produces a compact $b\bar b$ decay. The diagrams are schematic and do not establish completeness of the production calculation. The relative top- and bottom-quark contributions depend on their Yukawa couplings and loop form factors; their hierarchy cannot be inferred from a Yukawa ratio alone. Momentum-dependent scalar and pseudoscalar production with a recoil jet has been studied explicitly in Ref.~\cite{Field:2003pa}.

A quantitative prediction of the physical signal rate and hard-recoil spectrum requires a resolved loop calculation with the appropriate higher-order QCD corrections~\cite{Field:2003pa,KimWilliams:2024}. Resolving the loop dependence of the leading production amplitude and adding perturbative QCD corrections are separate ingredients. Both are beyond the scope of the present tagging study. Accordingly, the effective-vertex rates below serve as illustrative normalizations, while the main result concerns reconstruction of the squeezed heavy-flavour object.

\subsection{Event generation and parton-level topology}

Signal and background samples are generated at leading order with \texttt{MadGraph5\_aMC@NLO}~\cite{Alwall:2014hca}, using \texttt{NNPDF2.3} parton distributions~\cite{Ball:2013hta} in a four-flavour scheme. The signal includes one- and two-parton matrix-element samples. \texttt{PYTHIA8}~\cite{Bierlich:2022pfr} supplies showering and hadronization, with MLM merging~\cite{Mangano:2006rw} used to avoid counting the same resolved radiation in both the hard samples and the shower. The merging applies to the effective-vertex samples; it does not turn them into a resolved loop calculation. Detector response is simulated with \texttt{Delphes}~\cite{deFavereau:2013fsa}. The event-level rate illustration uses a centre-of-mass energy of 14 TeV.

The two matrix-element multiplicities describe different recoil configurations. With one accompanying parton, the transverse momentum of the pseudoscalar is balanced by a single hard parton at the matrix-element level. With two accompanying partons, that recoil can be shared between them, providing additional configurations for the candidate and the rest of the event. Including both multiplicities therefore samples the recoil geometry used in the subsequent event analysis. The parton shower supplies further radiation, and MLM merging combines its description with that of the hard samples. These ingredients connect production of a boosted light state to a reconstructed event containing a compact decay jet and accompanying recoil activity.

\begin{figure*}[t]
 \centering
 \includegraphics[width=0.485\textwidth]{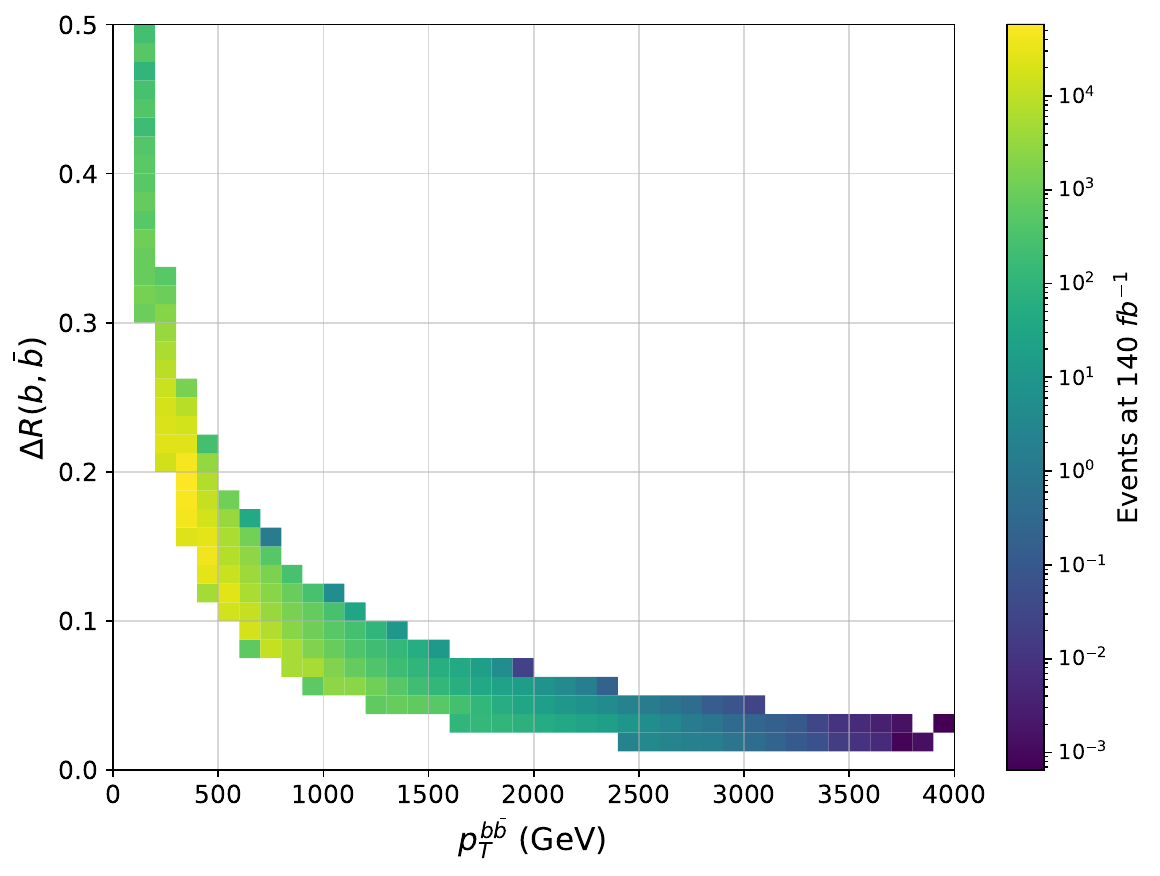}\hfill
 \includegraphics[width=0.485\textwidth]{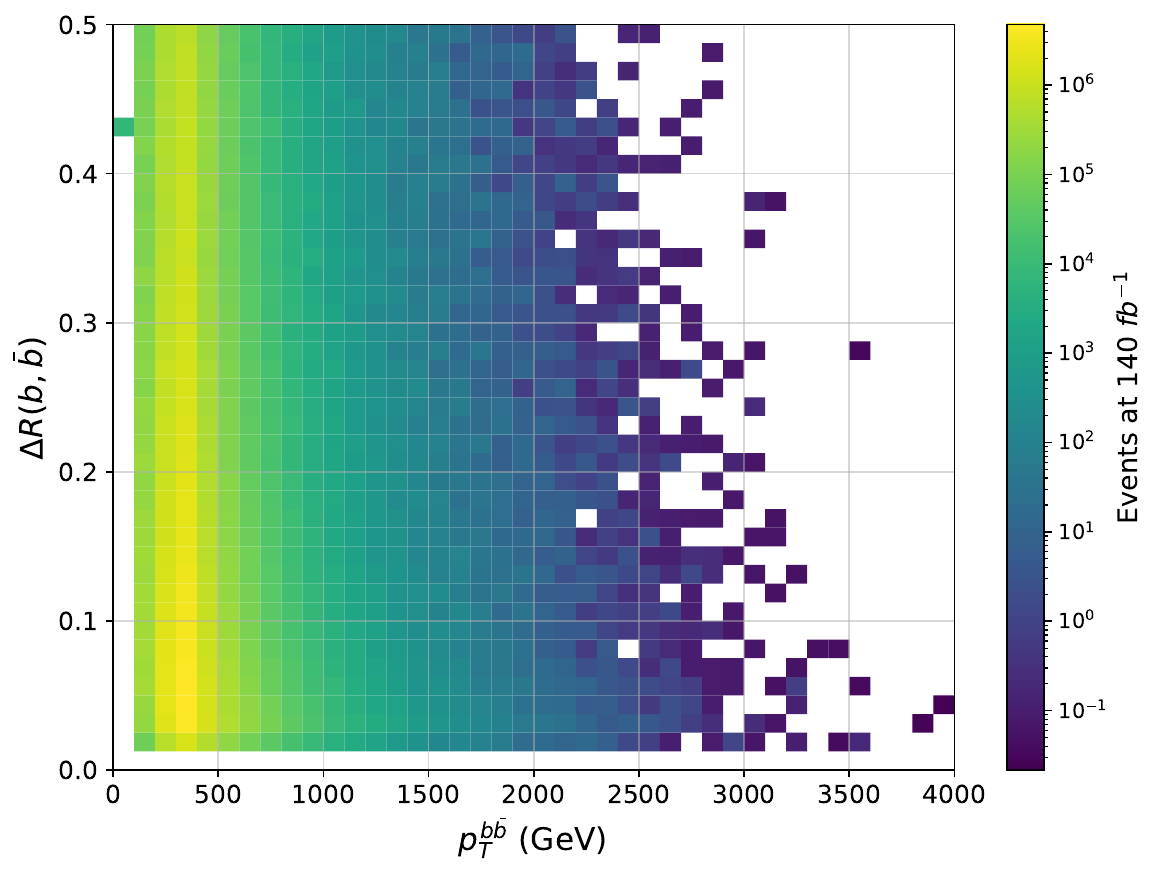}
 \caption{Parton-level distributions of $\Delta R(b,\bar b)$ versus $\pt^{b\bar b}$ for the $m_a=30$ GeV signal (BP1, left) and the QCD background (right). The signal shows decreasing angular separation with increasing transverse momentum, characteristic of the squeezed $b\bar b$ topology, while QCD populates a broader region. Colours represent expected event yields per bin at $140~\fb$, with separate logarithmic scales for the two panels.}
 \label{fig:parton_correlations}
\end{figure*}

The parton-level distributions in Fig.~\ref{fig:parton_correlations} connect the recoil requirement to the squeezed topology. In the BP1 signal panel, the $b\bar b$ pair originates from a parent with a fixed mass of 30 GeV. Increasing the pair transverse momentum is therefore accompanied by a decreasing angular separation, as expected from Eq.~\eqref{eq:opening}. The populated band follows this inverse relation, with a spread associated with asymmetric momentum sharing and departures from the ideal two-body approximation. The figure consequently shows how the production recoil controls the angular scale that the reconstruction must accommodate.

The QCD distribution occupies a broader region of the same plane. Its bottom-quark pairs are not constrained to reconstruct a common resonance mass, and the relation between their momentum and separation is correspondingly less localized. Nevertheless, gluon splitting also produces pairs at small $\Delta R$, including configurations in the angular region occupied by the signal. Requiring a compact pair thus selects a relevant topology without removing the QCD contribution. When both bottom quarks hadronize within one jet, this background also has genuine double-bottom content and must be separated using additional information, including the reconstructed mass and the event kinematics.

For the signal, the angular scale determines when the two decay systems can be captured by the common jet radius $R=0.5$. The relation in Eq.~\eqref{eq:opening} also explains why increasing the pseudoscalar mass at a given momentum makes containment more demanding. Conversely, a larger boost brings the decay products closer together while making two separate bottom jets harder to resolve. These observations motivate treating the compact pair as one reconstructed object and using its displaced tracks to identify the two bottom-hadron decay systems. Figure~\ref{fig:parton_correlations} supplies the parton-level motivation for this procedure; the detector-level event display and jet-category responses below describe the subsequent reconstruction and classification stages.

\subsection{Reconstructed jets and grooming}

\begin{figure*}[t]
 \centering
 \includegraphics[width=0.88\textwidth]{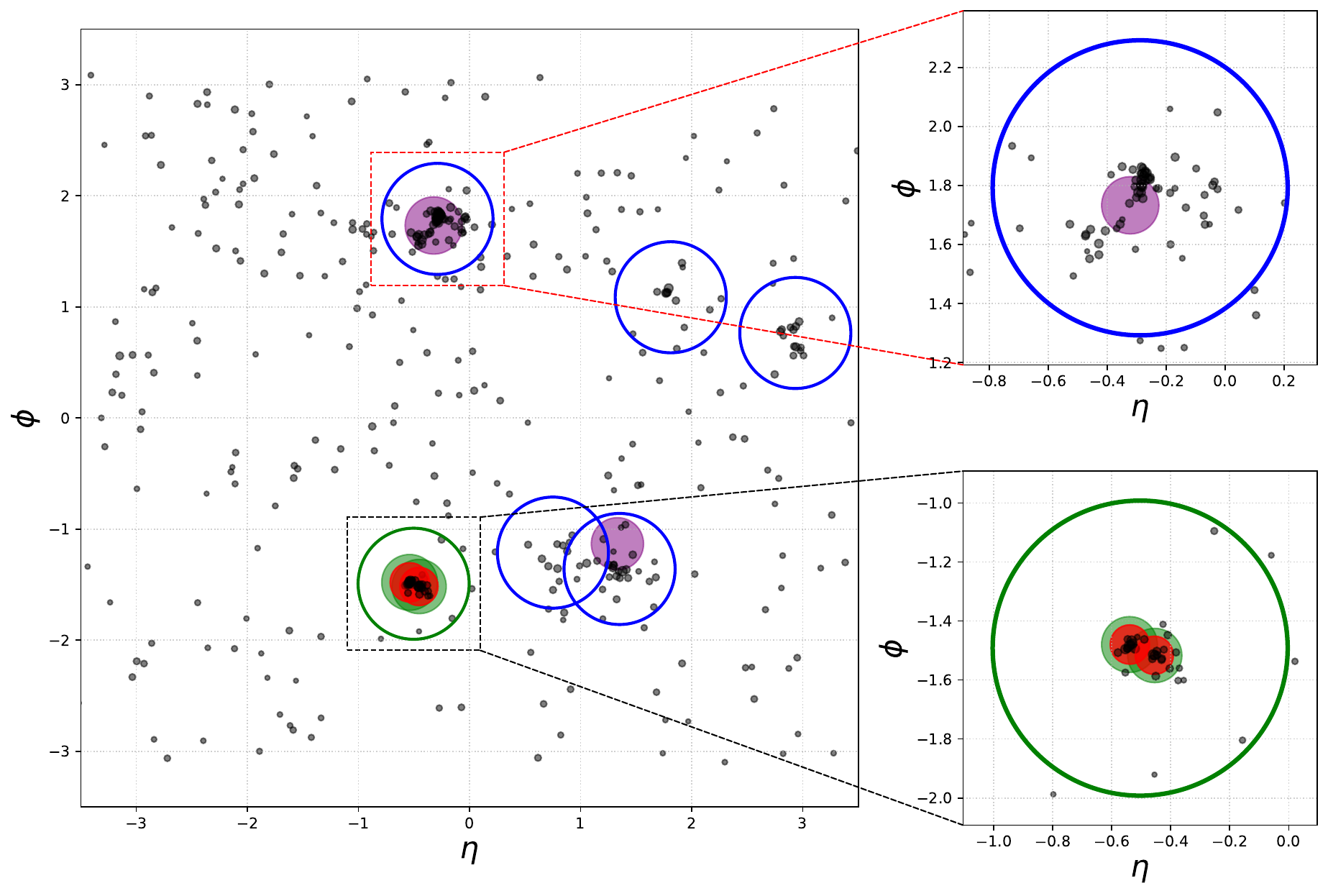}
 \caption{Signal-event display in the $\eta$--$\phi$ plane, illustrating the squeezed $b\bar b$ topology. Black points represent detector-level particles, red filled circles denote $B$ hadrons, and coloured filled circles indicate parton-level quarks and gluons. Unfilled circles show reconstructed jets with radius $R=0.5$. The lower inset highlights the two bottom-hadron decay systems contained within a single jet, while the upper inset shows a recoiling gluon and its associated radiation. Particle-marker sizes are scaled logarithmically with transverse momentum for visualization.}
 \label{fig:eta_phi_display}
\end{figure*}

Particle-flow objects are clustered with the anti-$k_t$ algorithm~\cite{Cacciari:2008gp}, using $R=0.5$. This radius captures the more collimated decays while limiting the angular region contributing unrelated activity. It is a common reconstruction choice for the three benchmarks, rather than an optimum established separately for each mass. We use soft drop to reduce soft, wide-angle contamination before computing the jet mass. The resulting mass is denoted by $\msd$. Figure~\ref{fig:eta_phi_display} illustrates this reconstruction in the $\eta$--$\phi$ plane, relating the parton directions and intermediate $B$ hadrons to the detector-level activity. In the lower inset, the two bottom quarks (green filled circles) and their associated $B$ hadrons (red filled circles) occupy a compact region inside a single reconstructed jet, indicated by the green unfilled circle. The black points show the nearby detector-level particles, whose overlapping activity is clustered into this candidate. Thus two heavy-flavour decay systems need not produce two resolved jets. The upper inset shows a recoiling gluon (purple filled circle) and the detector-level radiation within its reconstructed jet, outlined in blue. Such radiation structure also occurs in ordinary QCD jets, so the angular pattern alone does not identify a double-bottom decay. The displaced tracks associated with the two $B$ hadrons supply information about bottom-hadron multiplicity used by the tagger, even when their energy deposits belong to one jet. The event display provides a geometric illustration rather than a measurement of a tagging efficiency.

\section{Tracking-based squeezed double-\texorpdfstring{$b$}{b} tagging}
\label{sec:tagger}

\subsection{Jet categories and track observables}

The jet-level BDT distinguishes the $0b$, $1b$ and $2b$ categories. In the reference truth labelling these denote zero, one and two $B$ hadrons within the jet, respectively. The $0b$ label means non-bottom: it can include charm jets as well as light-quark and gluon jets. Truth labels are used to train and evaluate the classifier; reconstructed event selections use the tagger's output categories. The charm response is examined separately below.

The jet-level BDT uses jet kinematics, track and constituent multiplicities, charge information and track-displacement observables as inputs. The jet transverse momentum is denoted by $\pjet$. The track multiplicity $N_{\mathrm{trk}}$ counts the displaced tracks associated with the jet, whereas the constituent multiplicity $N_{\mathrm{const}}$ counts the reconstructed particle-flow objects clustered into it. The summed track charge $\sum q$ is the sum of the electric charges of the associated tracks. The quantities $N_{\mathrm{trk}}^+$ and $N_{\mathrm{trk}}^-$ count tracks with positive and negative charge, respectively. These observables describe the overall jet activity and charged-particle content.

The associated tracks also provide two-dimensional and three-dimensional impact parameters, denoted by $\ip_{2D}$ and $\ip_{3D}$. For each of these quantities, the displaced-track observables are evaluated in three intervals: below $100~\mu\mathrm m$, between $100~\mu\mathrm m$ and 10 mm, and above 10 mm. The first boundary is equivalently 0.1 mm. For an interval written as $a<\ip_D<b$, with $D=2D$ or $3D$, the notation $N_{\mathrm{trk}}(a<\ip_D<b)$ denotes the number of associated tracks whose impact parameters lie in that interval. The corresponding momentum fraction, $\frac{\sum p_T^{\mathrm{trk}}}{p_T^{\mathrm{jet}}}(a<\ip_D<b)$, is the scalar sum of the transverse momenta of these displaced tracks divided by the jet transverse momentum. The interval restriction applies to the tracks in the numerator. In particular, $\ndisplaced$ and $\fdisplaced$ quantify the displaced-track multiplicity and momentum fraction in the interval between 0.1 and 10 mm.

The $\pt$-weighted mean impact parameters are denoted by $\langle\ip_{2D}\rangle_{\pt}$ and $\langle\ip_{3D}\rangle_{\pt}$. Each is obtained by weighting the associated tracks' impact parameters by their transverse momenta and normalizing to the scalar sum of those track momenta. This normalization differs from that of the displaced-track momentum fraction, whose denominator is $\pjet$. The ordered inputs are $\ip_{2D}^{(r)}$ and $\ip_{3D}^{(r)}$, for $r=1,\ldots,5$. Tracks are sorted separately by decreasing impact-parameter magnitude in each dimension. Thus $\ip_{2D}^{(5)}$ is the fifth-largest 2D impact parameter, while $\ip_{3D}^{(3)}$ is the third-largest 3D impact parameter; the superscript does not denote a momentum rank. The feature set combines these ordered values and weighted means with the jet, charge and interval-based observables defined above. Together they describe the distribution of displaced activity across the jet. The groomed mass is not among these jet-classification inputs.

\subsection{Training and discriminating information}

The simulated-jet dataset is randomly divided into 70\% training, 15\% validation and 15\% test samples. A multiclass \texttt{XGBoost} BDT is trained with early stopping after twenty rounds without validation improvement. Table~\ref{tab:btagger_params} gives the training configuration. The response probabilities quoted below refer to this reference simulation and its jet mixture; they should not be read as efficiencies independent of jet momentum or detector conditions.

\begin{table}[t]
 \caption{Training configuration of the jet-level BDT used to classify jets into the $0b$, $1b$ and $2b$ categories.}
 \label{tab:btagger_params}
 \centering
 \begin{ruledtabular}
 \begin{tabular}{lc}
 Objective & \texttt{multi:softprob}\\
 Number of classes & 3\\
 Maximum estimators & 500\\
 Learning rate & 0.015\\
 Maximum depth & 2\\
 Minimum child weight & 2\\
 Subsample fraction & 0.8\\
 Column fraction per tree & 1\\
 Evaluation metric & \texttt{mlogloss}\\
 Early-stopping rounds & 20
 \end{tabular}
 \end{ruledtabular}
\end{table}

\begin{figure}[t]
 \centering
 \includegraphics[width=\columnwidth]{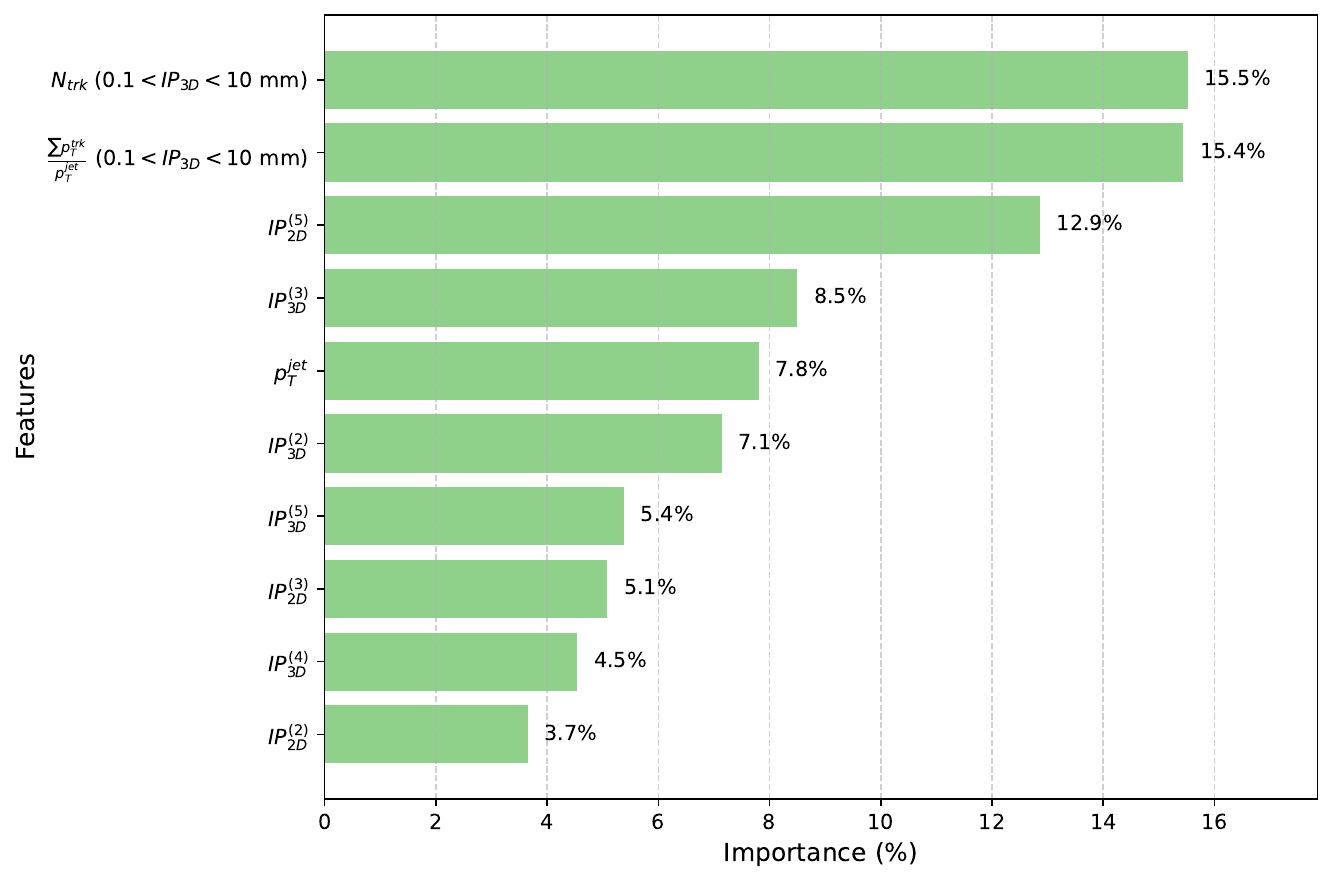}
 \caption{Gain-based ranking of the leading input features used by the jet-level BDT to distinguish $0b$, $1b$ and $2b$ jets. Displaced-track multiplicities, momentum fractions and ordered impact parameters provide the dominant discrimination.}
 \label{fig:btagger_importance}
\end{figure}

Figure~\ref{fig:btagger_importance} shows that the most useful inputs are $\ndisplaced$ and $\fdisplaced$, followed by ordered impact parameters such as $\ip_{2D}^{(5)}$ and $\ip_{3D}^{(3)}$. This preference for track-displacement observables reflects the two bottom-hadron decays within a squeezed $2b$ jet. Their decay products provide a larger population of displaced tracks than is generally present in a jet containing one or no bottom hadron. The resulting track pattern retains information about the two-bottom configuration even when the decay products are clustered into a single jet.

The displaced-track count and momentum fraction characterize complementary aspects of this pattern. The count measures how many tracks populate the specified displacement interval, while the momentum fraction measures their contribution to the jet transverse momentum. Their joint use therefore describes both the multiplicity and momentum carried by displaced tracks. The importance of $\ip_{2D}^{(5)}$ and $\ip_{3D}^{(3)}$ further shows that the classifier uses displacements across several tracks. These ordered values distinguish a population of displaced tracks from a configuration characterized mainly by one large impact parameter. The jet momentum $\pjet$ and the remaining ordered displacement inputs supplement this information with the jet kinematics and the wider displacement pattern.

\begin{figure*}[t]
 \centering
 \includegraphics[width=0.96\textwidth]{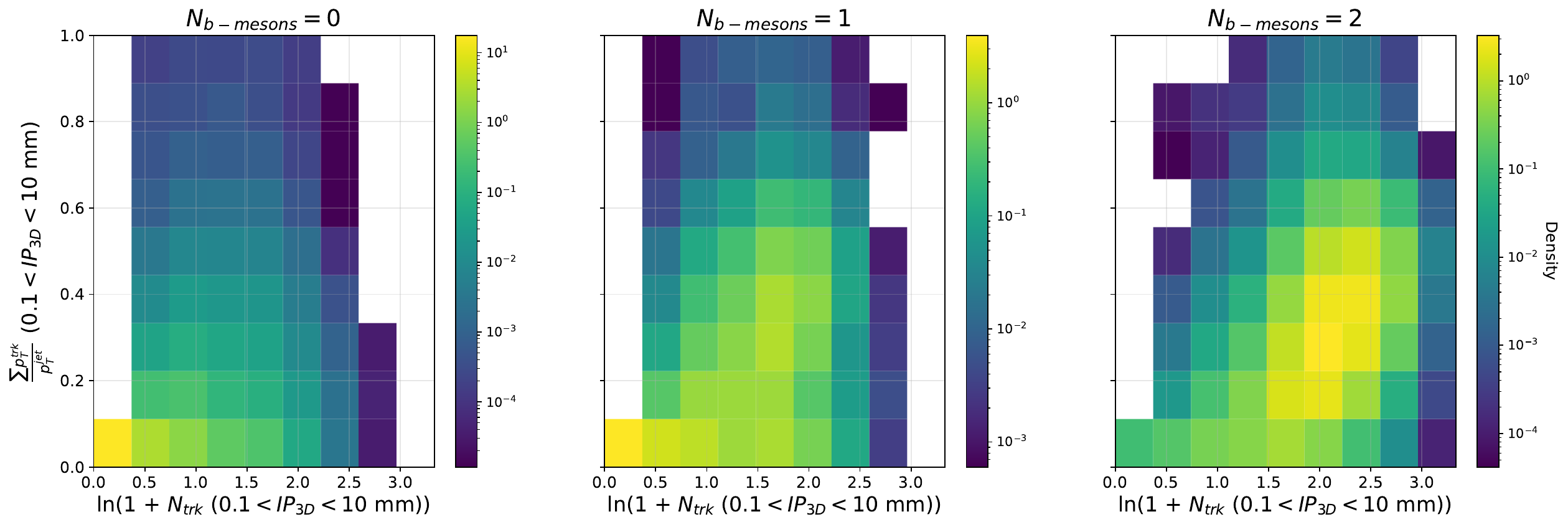}\\[-1mm]
 \includegraphics[width=0.96\textwidth]{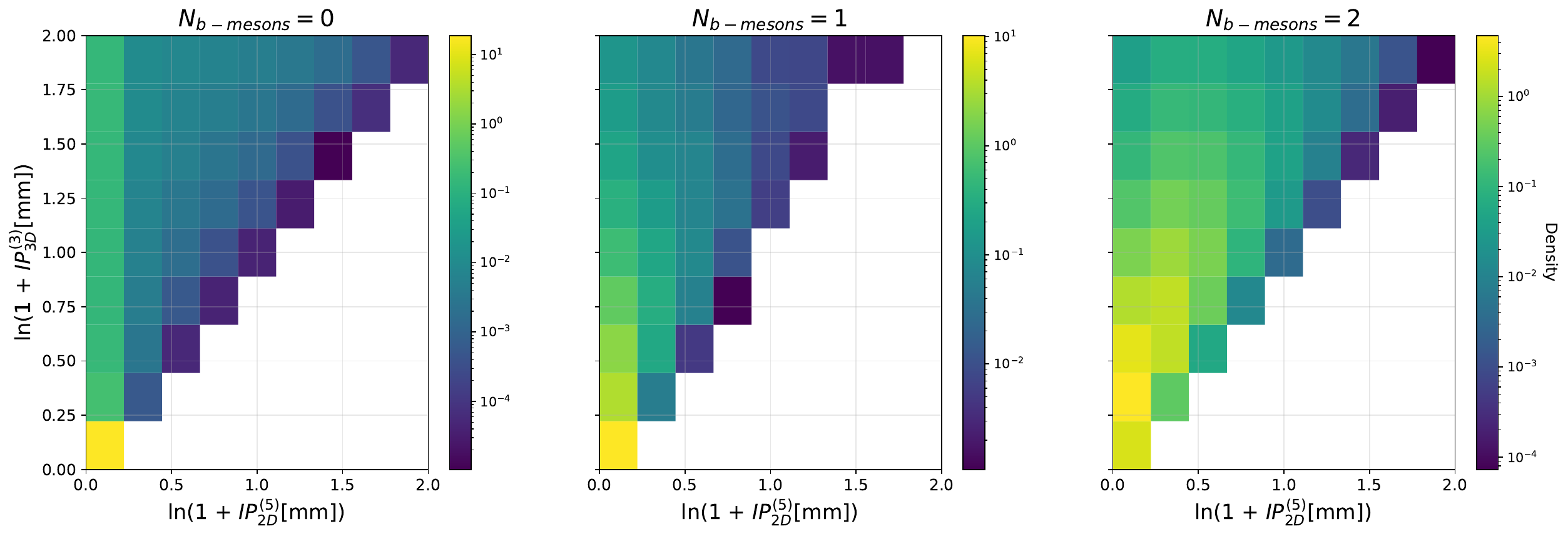}\\[-1mm]
 \includegraphics[width=0.96\textwidth]{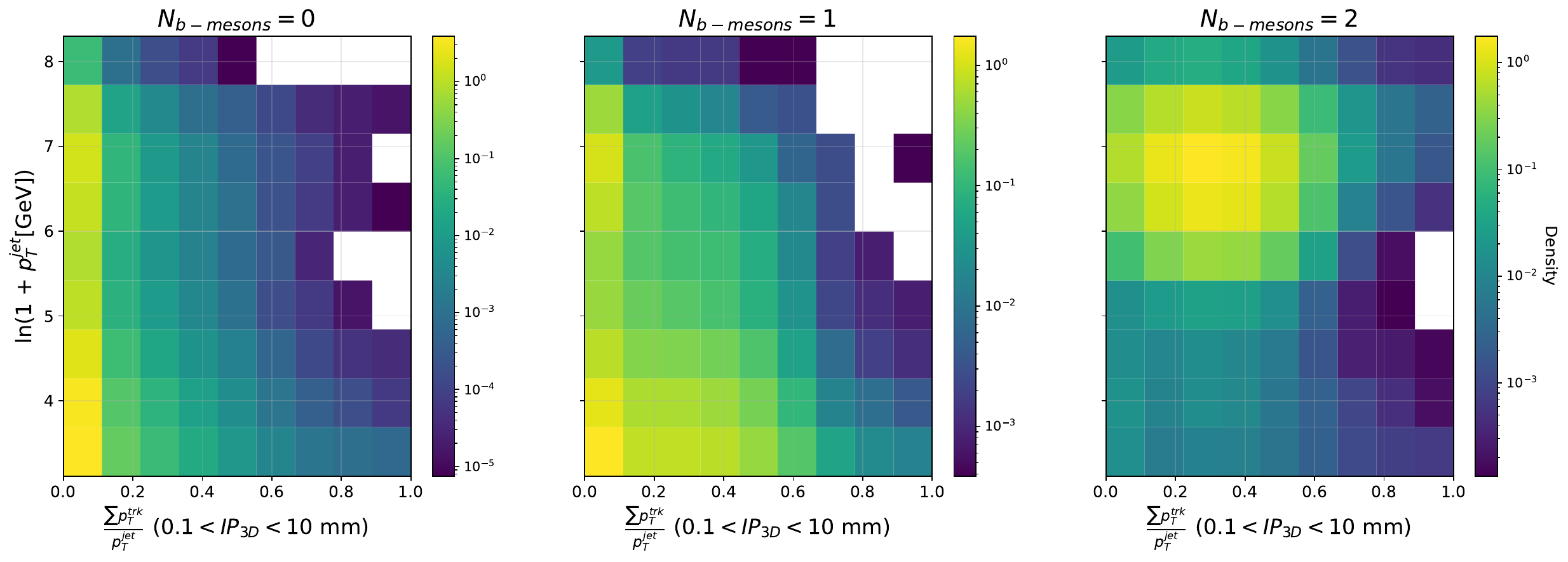}
 \caption{Tracking-variable distributions for true $0b$ (left), $1b$ (middle) and $2b$ (right) jets. Top row: $\ndisplaced$ and $\fdisplaced$. Middle row: the fifth-largest 2D impact parameter $\ip_{2D}^{(5)}$ and third-largest 3D impact parameter $\ip_{3D}^{(3)}$. Bottom row: the same displaced-track momentum fraction and the jet transverse momentum $\pjet$. Each panel has its own logarithmic density scale. The distributions show increasing displaced activity with bottom-hadron multiplicity, together with substantial overlap between categories.}
 \label{fig:flavor_ip_correlations}
\end{figure*}

The distributions in Fig.~\ref{fig:flavor_ip_correlations} illustrate how these inputs distinguish the $0b$, $1b$ and $2b$ categories. The top row compares the displaced-track multiplicity with the corresponding momentum fraction. The $2b$ population extends towards both larger counts and larger momentum fractions, as expected when two bottom-hadron decays contribute tracks to the same jet. The $0b$ distribution is concentrated towards smaller displaced activity, while the $1b$ population overlaps the other two categories. In the middle row, $\ip_{2D}^{(5)}$ and $\ip_{3D}^{(3)}$ probe whether sizeable displacements persist to the fifth and third entries of the ordered track lists. The concentration near small values for $0b$ jets contrasts with the broader displaced populations in the bottom-containing categories.

The bottom row relates the displaced-track momentum fraction to $\pjet$. Over the displayed momentum range, $2b$ jets retain a substantial contribution from displaced tracks, whereas the non-bottom population is concentrated at smaller fractions. The $1b$ category again occupies an overlapping region, emphasizing the need to distinguish two bottom hadrons from one. This momentum dependence connects the tracking observables to the kinematic input used by the BDT. Together, the three rows exhibit the displaced-track multiplicity, momentum sharing and ordered displacements underlying the feature ranking in Fig.~\ref{fig:btagger_importance}. The separate colour scales preclude interpreting visual colour intensity as the same rate in all three columns.

\subsection{Bottom and charm response}

Tables~\ref{tab:conf_matrix_bJet} and \ref{tab:conf_matrix_cJet} give the reported category responses. Approximately $83\%$ of true double-bottom jets enter the tagged $2b$ category. The corresponding responses for true $1b$ and $0b$ jets are $9.7\%$ and $0.5\%$, respectively. The single-bottom leakage is therefore appreciable even when the non-bottom rejection is strong. Its contribution to an event sample depends on the much larger input rates of the different QCD categories.

\begin{table}[t]
 \caption{Confusion matrix of the jet-level BDT for the $0b$, $1b$ and $2b$ categories. Rows indicate the true bottom-hadron multiplicity, while columns indicate the category assigned by the classifier.}
 \label{tab:conf_matrix_bJet}
 \centering
 \begin{ruledtabular}
 \begin{tabular}{lccc}
 True jet & Tagged $0b$ & Tagged $1b$ & Tagged $2b$\\
 $0b$ & 0.90 & 0.08 & 0.005\\
 $1b$ & 0.12 & 0.78 & 0.097\\
 $2b$ & 0.016 & 0.15 & 0.83
 \end{tabular}
 \end{ruledtabular}
\end{table}

\begin{table}[t]
 \caption{Response of jets containing zero, one or two charm hadrons under the jet-level BDT. Rows indicate the true charm multiplicity, while columns indicate the assigned $0b$, $1b$ and $2b$ categories.}
 \label{tab:conf_matrix_cJet}
 \centering
 \begin{ruledtabular}
 \begin{tabular}{lccc}
 True jet & Tagged $0b$ & Tagged $1b$ & Tagged $2b$\\
 $0c$ & 0.93 & 0.06 & 0.001\\
 $1c$ & 0.37 & 0.60 & 0.017\\
 $2c$ & 0.14 & 0.74 & 0.11
 \end{tabular}
 \end{ruledtabular}
\end{table}

The off-diagonal entries in Table~\ref{tab:conf_matrix_bJet} describe migration between the reconstructed jet categories. Alongside the $83\%$ response in the tagged $2b$ category, true $2b$ jets have reported responses of $15\%$ in the $1b$ category and $1.6\%$ in the $0b$ category. Thus the larger migration for true double-bottom jets is into the single-bottom category. Conversely, the $9.7\%$ and $0.5\%$ responses of true $1b$ and $0b$ jets in the tagged $2b$ category contribute to the selected background. Charm must also be considered separately: Table~\ref{tab:conf_matrix_cJet} gives $2b$ responses of $1.7\%$ for $1c$ jets and $11\%$ for $2c$ jets. The larger response of the latter category reflects the difficulty of separating a jet with two charm hadrons from the two-bottom topology using displaced tracks. These responses must be considered together with the input populations of the categories, since a small mistag rate can contribute appreciably when the corresponding background yield is large. The displayed matrices summarize the reference samples; the numerical entries are retained without imposing an additional row renormalization.

\subsection{Groomed-mass response}

\begin{figure*}[t]
 \centering
 \includegraphics[width=0.485\textwidth]{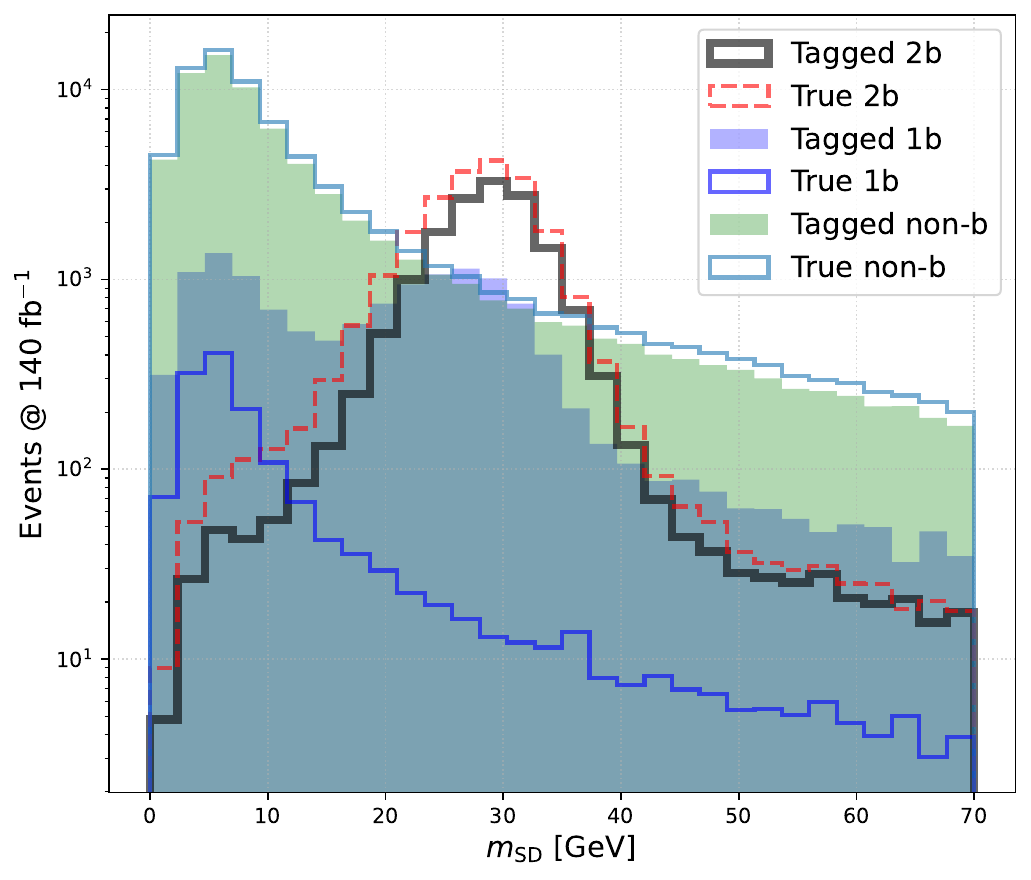}\hfill
 \includegraphics[width=0.485\textwidth]{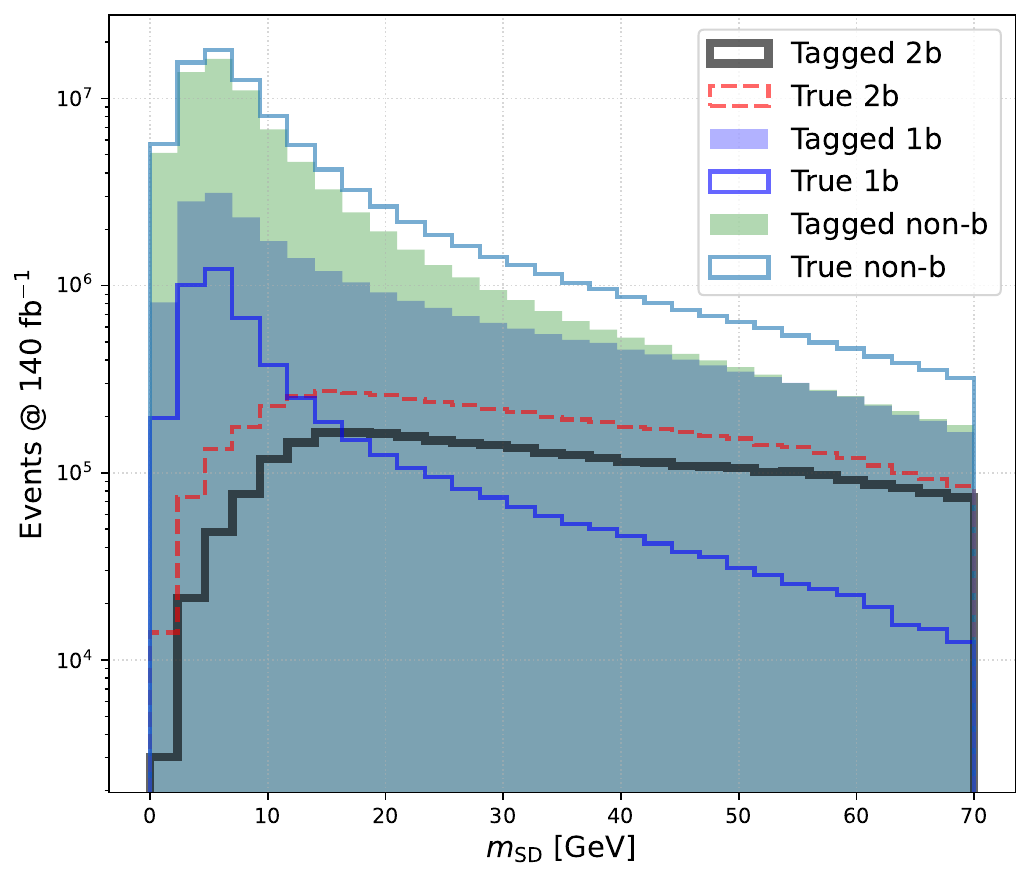}
 \caption{Soft-drop mass distributions for the $m_a=30$ GeV signal (BP1, left) and the QCD background (right), normalized to $140~\fb$. Jets classified by their true bottom-hadron content are compared with those assigned by the BDT to the $0b$, $1b$ and $2b$ categories. The tagged $2b$ signal retains a peak near the pseudoscalar mass, while the corresponding QCD distribution forms a continuum.}
 \label{fig:sd_mass_dist}
\end{figure*}

Figure~\ref{fig:sd_mass_dist} compares the reconstructed soft-drop mass spectra before and after classification into the $0b$, $1b$ and $2b$ jet categories. In the BP1 signal panel, the true $2b$ spectrum (red dashed line) has a localized peak near $m_a=30$ GeV. The tagged $2b$ spectrum (thick grey line) retains this peak and approximately follows its shape, showing that the squeezed decay remains identifiable after clustering, grooming and tagging. Both spectra extend beyond the peak, reflecting the spread in the reconstructed jet mass. The non-bottom jets are concentrated at lower masses and form a falling distribution, while the tagged $1b$ spectrum also contains activity in the resonance region. The latter is consistent with migration between the $0b$, $1b$ and $2b$ jet categories. The signal panel includes other jets in the event as well as the resonance decay, so its inclusive jet population need not be dominated by true $2b$ jets. The overlaid spectra compare category yields; their ratio is not a direct measurement of the tagging efficiency.

In the QCD panel, non-bottom jets form the largest population before tagging. The tagged $2b$ distribution is substantially reduced relative to that population, but remains a sizeable continuum. Its broad low-mass turn-on and falling higher-mass spectrum show no localized structure near 30 GeV comparable to the signal peak. This difference motivates combining the double-$b$ tag with a groomed-mass window: tagging reduces the non-bottom contribution, while the mass selection targets the resonance within the remaining continuum.

The tagged QCD $2b$ spectrum receives contributions from genuine compact $b\bar b$ systems and from category migration. Schematically, after integrating over the other jet kinematics,
\begin{equation}
 \frac{dN_{\mathrm{tag}\,2b}}{d\msd}
 =\sum_c\int dx\,
 \epsilon_{c\to2b}(\msd,x)
 \frac{d^2N_c}{d\msd\,dx},
 \label{eq:mass-response}
\end{equation}
Here $N_{\mathrm{tag}\,2b}$ is the expected number of reconstructed jets tagged as $2b$, and $dN_{\mathrm{tag}\,2b}/d\msd$ is their distribution in reconstructed soft-drop mass $\msd$. The index $c$ labels the mutually exclusive truth categories $0b$, $1b$ and $2b$ according to their bottom-hadron content; $N_c$ is the corresponding jet yield before jet classification. The variable $x$ denotes other jet kinematics, such as transverse momentum and pseudorapidity, and $d^2N_c/(d\msd\,dx)$ is the pretag yield per unit mass and kinematic interval. The probability $\epsilon_{c\to2b}(\msd,x)$ describes assignment of a jet from truth category $c$ to the tagged $2b$ category at that mass and kinematics: it is an efficiency for $c=2b$ and a mistag probability otherwise. The integral $\int dx$ covers the accepted kinematics, including all components if $x$ represents several variables, while $\sum_c$ adds the truth-category contributions. Each pretag distribution is therefore weighted by its tagging probability.

A small mistag probability can still matter when the corresponding pretag yield $N_c$ is large. The background spectrum in the mass region of interest is smooth, while the signal has a localized mass structure. This motivates using $\msd$ in the event analysis. Since displacement and momentum inputs can correlate with mass, a smooth distribution in this simulation does not establish a universal absence of tag-induced mass distortion.

\section{Event-level search strategy}
\label{sec:event}

\subsection{Backgrounds}

QCD multijets are the dominant background. Genuine $g\to b\bar b$ splittings can put two bottom hadrons in the same jet and supply an irreducible heavy-flavour component. Reducible contributions arise from single-bottom, charm, light-quark and gluon jets assigned to the $2b$ category. Both components are included in the interpretation of the tagged background; their relative importance depends on the input sample mixture and the kinematic selection.

Other potential backgrounds include $W/Z$ production with jets, $t\bar t$, $t\bar tV$, diboson and triboson production. To select the collimated $b\bar b$ configurations relevant to a light resonance, the parton-level background samples are generated with the angular restriction
\begin{equation}
 0.02<\Delta R(b,\bar b)<0.9.
 \label{eq:background-generation}
\end{equation}
This restriction, together with the subsequent squeezed-jet selection, suppresses the backgrounds from top quarks and massive vector bosons, whose decay products typically have larger angular separations or additional resolved activity. Their remaining contributions are negligible compared with QCD in the samples considered here.

To account approximately for higher-order QCD corrections, the leading-order QCD background cross section is scaled by a $K$-factor of 1.3. Here $K$ represents the ratio of next-to-leading-order to leading-order cross sections~\cite{Kim:2024ppt}.

\subsection{Jet notation and preselection}

After classification into the $0b$, $1b$ and $2b$ categories, jets in each category are ordered by decreasing transverse momentum. We denote the $i$th jet in category $cb$ by $J_i^{cb}$, with $c=0,1,2$. Thus $\jbb$ is the leading tagged double-bottom jet, $J_1^{1b}$ is the leading single-bottom jet, and $J_1^{0b}$ and $J_2^{0b}$ are the leading and second-hardest non-bottom jets. The labels describe reconstructed categories, not the truth content of the selected event. The multiplicities $N_{cb}$ count jets in each category above the analysis threshold. The reconstructed-jet transverse momenta are denoted by $P_T(J_i^{cb})$, where $P_T$ is equivalent to the transverse momentum $p_T$ used elsewhere.

The common preselection is
\begin{equation}
 \begin{gathered}
 N_{2b}=1,\qquad N_{0b}+N_{1b}\geq1,\\
 P_T(J)>100~\GeV\quad\text{for analysis jets}.
 \end{gathered}
 \label{eq:preselection}
\end{equation}
The single $2b$ candidate is accompanied by at least one recoil jet. The same momentum threshold is used for all three masses to make the benchmark comparison at a common baseline. Equation~\eqref{eq:opening} shows why a mass-dependent recoil requirement could change containment and mass resolution; the present study does not claim that the common threshold is optimal for every mass.

The candidate is then required to lie in a benchmark-dependent groomed-mass interval:
\begin{equation}
 \begin{array}{ll}
 15\leq\msd(\jbb)\leq45~\GeV,&\mathrm{BP1},\\
 35\leq\msd(\jbb)\leq65~\GeV,&\mathrm{BP2},\\
 45\leq\msd(\jbb)\leq75~\GeV,&\mathrm{BP3}.
 \end{array}
 \label{eq:mass-windows}
\end{equation}
These windows use the reconstructed mass after tagging. They retain the signal region while reducing the broad QCD continuum, before the final event-level discrimination.

\subsection{Event classifier}

A separate event BDT is trained for each benchmark, using 80\% of the event dataset for training, 10\% for validation and 10\% for testing. Its configuration is given in Table~\ref{tab:evt_bdt_params}. The event classifier uses 97 inputs describing global activity, the tagged candidate and correlations with recoil jets. The input groups comprise jet multiplicities, missing transverse energy, individual jet momenta and energies, the candidate's groomed mass and substructure, pairwise angular separations, and invariant masses formed from the tagged jets. The multiplicities $N_{0b}$ and $N_{1b}$ count the selected non-bottom and single-bottom jets, respectively. The missing transverse energy is denoted by $E_T^{\mathrm{miss}}$.

\begin{table}[t]
 \caption{Hyperparameters used to train the event-level BDT for discriminating signal events from the Standard Model background.}
 \label{tab:evt_bdt_params}
 \centering
 \begin{ruledtabular}
 \begin{tabular}{lc}
 Maximum estimators & 300\\
 Learning rate & 0.01\\
 Maximum depth & 3\\
 Minimum child weight & 2\\
 Subsample fraction & 0.8\\
 Reported evaluation metric & \texttt{mlogloss}\\
 Early-stopping rounds & 20
 \end{tabular}
 \end{ruledtabular}
\end{table}

The momentum inputs include $P_T(\jbb)$, $P_T(J_1^{1b})$, $P_T(J_1^{0b})$ and $P_T(J_2^{0b})$. These are the transverse momenta of the leading double-bottom candidate, the leading single-bottom jet, and the two leading non-bottom jets, respectively. The quantity $E(J_1^{1b})$ is the energy of the leading single-bottom jet. For the candidate, $\msd(\jbb)$ denotes the soft-drop mass. The substructure ratios $\tau_{21}(\jbb)=\tau_2(\jbb)/\tau_1(\jbb)$ and $\tau_{32}(\jbb)=\tau_3(\jbb)/\tau_2(\jbb)$ compare its $N$-subjettiness values~\cite{Thaler:2010tr}, which quantify how the jet constituents are distributed relative to $N$ candidate subjet axes.

For two reconstructed jets, $\Delta R(J_a,J_b)$ denotes their separation in the $\eta$--$\phi$ plane, while $\Delta\Phi(J_a,J_b)$ denotes their azimuthal separation. The invariant mass of the jet pair is written as $m(J_a,J_b)$ and is formed from their summed four-momenta. The angular inputs include $\Delta R(J_1^{0b},J_2^{0b})$, $\Delta\Phi(J_1^{0b},\jbb)$ and $\Delta R(J_1^{1b},\jbb)$. The pair-mass inputs include $m(J_2^{0b},\jbb)$, $m(J_1^{0b},J_2^{0b})$ and $m(J_1^{1b},\jbb)$. These quantities describe the recoil geometry and the mass scales of combinations involving the candidate and other tagged jets.

\begin{figure}[t]
 \centering
 \includegraphics[width=\columnwidth]{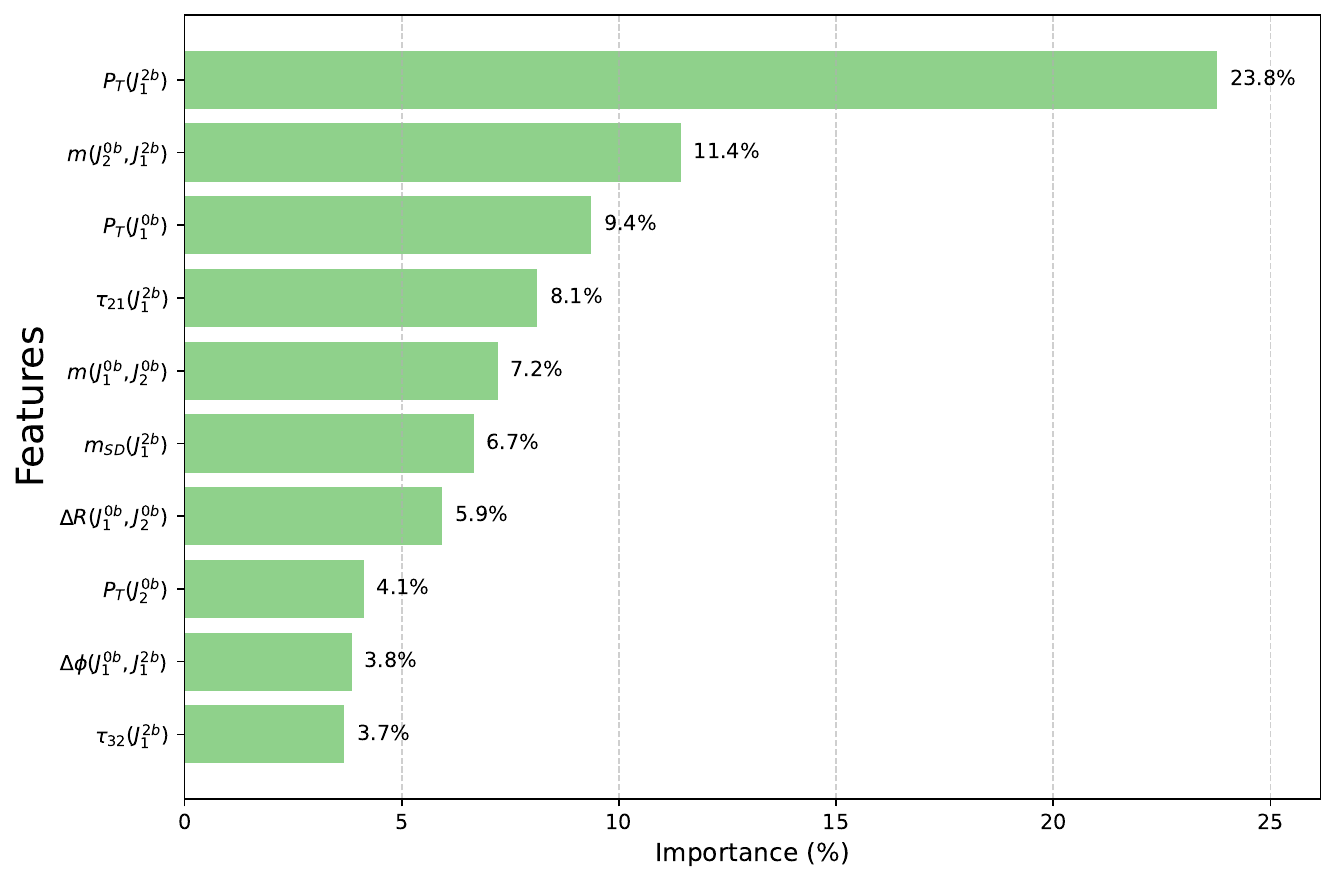}
 \caption{Relative importance of the leading input features used by the event-level BDT for the $m_a=30$ GeV benchmark (BP1). The transverse momentum of the squeezed double-$b$ candidate and its pair invariant mass with a recoil jet are the highest-ranked inputs for signal--background discrimination.}
 \label{fig:feature_importance}
\end{figure}

Figure~\ref{fig:feature_importance} shows that the event classifier uses information beyond the jet-category label. The leading input is $P_T(\jbb)$, followed by the pair mass $m(J_2^{0b},\jbb)$. The candidate transverse momentum characterizes the boost of the squeezed $b\bar b$ system, while the momenta of the other jets describe the accompanying hard recoil. The pair masses combine the kinematics of these objects: $m(J_2^{0b},\jbb)$ and $m(J_1^{1b},\jbb)$ involve the candidate and another jet, whereas $m(J_1^{0b},J_2^{0b})$ characterizes a pair in the recoil system. Together with the individual jet momenta, these masses describe the momentum configuration of the event. The $\Delta R$ and $\Delta\Phi$ inputs supplement this information by recording the relative geometry of the candidate and the other reconstructed jets.

The candidate's groomed mass and substructure ratios provide complementary information about its internal structure. The soft-drop mass follows the reconstructed resonance scale, while $\tau_{21}(\jbb)$ characterizes the two-prong structure expected from a collimated $b\bar b$ decay. The ratio $\tau_{32}(\jbb)$ supplies a further comparison of the constituent distribution relative to different numbers of subjet axes. The BDT combines these candidate properties with the momentum, pair-mass and angular inputs, allowing correlations between the reconstructed decay and the recoil topology to contribute to the signal--background separation after the pre-selection and mass-window requirements.

This is a benchmark-dependent search stage, with a separate event classifier trained for each pseudoscalar mass. Its performance is therefore kept distinct from the jet tagger's category response, which is the reusable reconstruction component. The event BDT uses the production kinematics and reconstructed candidate properties to reduce the background remaining after the squeezed-$2b$ selection.

\section{Rate illustration and statistical sensitivity}
\label{sec:sensitivity}

\subsection{Reference normalization and event yields}

Table~\ref{tab:yields} gives the reference yields at 14 TeV and $140~\fb$. All numerical entries use the effective-vertex signal normalization. The input row corresponds to baseline sample rates of 0.72, 0.34 and 0.25 pb for BP1, BP2 and BP3, and approximately $1.63\times10^3$ pb for the combined background. These numbers follow from the input yields divided by luminosity. They characterize the samples entering the cut flow and are not inclusive, resolved-loop production cross sections.

\begin{table*}[t]
 \caption{Expected event yields for the three signal benchmarks and the QCD background after successive selection requirements at $\sqrt{s}=14$ TeV and an integrated luminosity of $140~\fb$. Background yields include a $K$-factor of 1.3. K and M denote $10^3$ and $10^6$ events, respectively, and mass windows are given in GeV.}
 \label{tab:yields}
 \centering
 \begin{ruledtabular}
 \begin{tabular}{lrrrr}
 Selection & BP1 & BP2 & BP3 & Background\\
 Input sample & 100.8 K & 47.6 K & 35.0 K & 228.57 M\\
 Preselection, Eq.~\eqref{eq:preselection} & 70.47 K & 33.6 K & 24.73 K & 27.91 M\\
 BP1 mass window: $15\leq\msd(\jbb)\leq45$ & 68.6 K & --- & --- & 11.85 M\\
 BP2 mass window: $35\leq\msd(\jbb)\leq65$ & --- & 30.8 K & --- & 9.38 M\\
 BP3 mass window: $45\leq\msd(\jbb)\leq75$ & --- & --- & 21.0 K & 8.31 M\\
 BP1 event-BDT score $>0.87$ & 1.37 K & --- & --- & 1.05 K\\
 BP2 event-BDT score $>0.87$ & --- & 0.50 K & --- & 0.67 K\\
 BP3 event-BDT score $>0.88$ & --- & --- & 0.23 K & 0.41 K
 \end{tabular}
 \end{ruledtabular}
\end{table*}

The successive rows of Table~\ref{tab:yields} show how the tagged topology, reconstructed mass and event kinematics reduce the background. The common preselection retains 70.47 K, 33.6 K and 24.73 K signal events for BP1, BP2 and BP3, respectively. These correspond to approximately $69.9\%$, $70.6\%$ and $70.7\%$ of their input yields. Over the same stage, the background falls from 228.57 M to 27.91 M events, retaining approximately $12.2\%$. The preselection therefore establishes a candidate-plus-recoil sample with a larger signal fraction, while leaving a substantial QCD population for the subsequent discrimination. These fractions refer to the combined event requirements in Eq.~\eqref{eq:preselection}, including the tagged-jet multiplicities and momentum threshold.

The benchmark-dependent mass windows then select the reconstructed resonance region. For BP1, the interval from 15 to 45 GeV retains 68.6 K of the 70.47 K preselected signal events, while the background decreases to 11.85 M. The corresponding signal yields for BP2 and BP3 are 30.8 K and 21.0 K, with backgrounds of 9.38 M and 8.31 M in their respective windows. Relative to the preselected samples, the signal retentions are approximately $97.3\%$, $91.7\%$ and $84.9\%$, while the background retentions are $42.5\%$, $33.6\%$ and $29.8\%$. These differences describe the selected samples and reconstructed mass intervals for the three benchmarks. The resulting yields show the usefulness of a localized signal mass structure within the broader QCD continuum.

Even after the mass selection, the background remains much larger than the signal. The event BDT then uses the candidate and recoil information described in Sec.~\ref{sec:event}. At the reported score thresholds, the selected signal and background yields are 1.37 K and 1.05 K for BP1, 0.50 K and 0.67 K for BP2, and 0.23 K and 0.41 K for BP3. The final ratios $S/B$ are consequently about 1.30, 0.75 and 0.56. Relative to the mass-window samples, the event-BDT selection retains approximately $2.0\%$, $1.6\%$ and $1.1\%$ of the signal, respectively, together with less than $0.01\%$ of the background in each case. The substantial background reduction is thus accompanied by a restrictive signal selection, as is relevant when interpreting the final sensitivity.

All percentages and ratios in this discussion are calculated from the rounded yields in Table~\ref{tab:yields}. The retention percentages use the samples entering the corresponding selection stage as their denominators. They describe event-level retention and should not be identified with the true-$2b$ jet response in Table~\ref{tab:conf_matrix_bJet}. The cut flow illustrates the combined use of double-$b$ tagging, a groomed-mass interval and recoil correlations. The final thresholds of 0.87, 0.87 and 0.88 belong to the event BDT and should not be confused with a jet-category probability.

For a fixed signal distribution, one may replace the reference input rate by $r\sigma_{\mathrm{ref}}$. The selected signal yield then becomes $rS_{\mathrm{ref}}$, while the background and the efficiencies for that distribution are held fixed. This gives a direct rate interpretation through Eq.~\eqref{eq:yield-factorization}. It does not provide the coupling dependence of a physical production cross section; such a dependence requires the production calculation and the decay normalization.

\subsection{Significance with background uncertainty}

For selected signal and background counts $S$ and $B$, we use the asymptotic profile-likelihood significance~\cite{Cowan:2010js} with a background-normalization uncertainty $\sigma_B=\epsilon B$. Here $\sigma_B$ is the absolute uncertainty in the expected background yield, and $\epsilon=\sigma_B/B$ is its fractional systematic uncertainty:
\begin{align}
 Z_A^2=2\Bigg[&(S+B)\ln\left(1+\frac{S}{B+\epsilon^2B(S+B)}\right)
 \nonumber\\[-1mm]
 &-\frac{1}{\epsilon^2}\ln\left(1+\frac{\epsilon^2S}{1+\epsilon^2B}\right)\Bigg].
 \label{eq:significance}
\end{align}
We take $\epsilon=0.1$, corresponding to a 10\% systematic uncertainty in the background yield, for the reference sensitivity. The luminosity curves assume that the selected signal and background rates scale linearly with $\mathcal L$, while the fractional background uncertainty remains fixed.

\begin{figure}[t]
 \centering
 % Native vector overlays correct only the two textual labels; curves are unchanged.
 \begin{overpic}[width=\columnwidth]{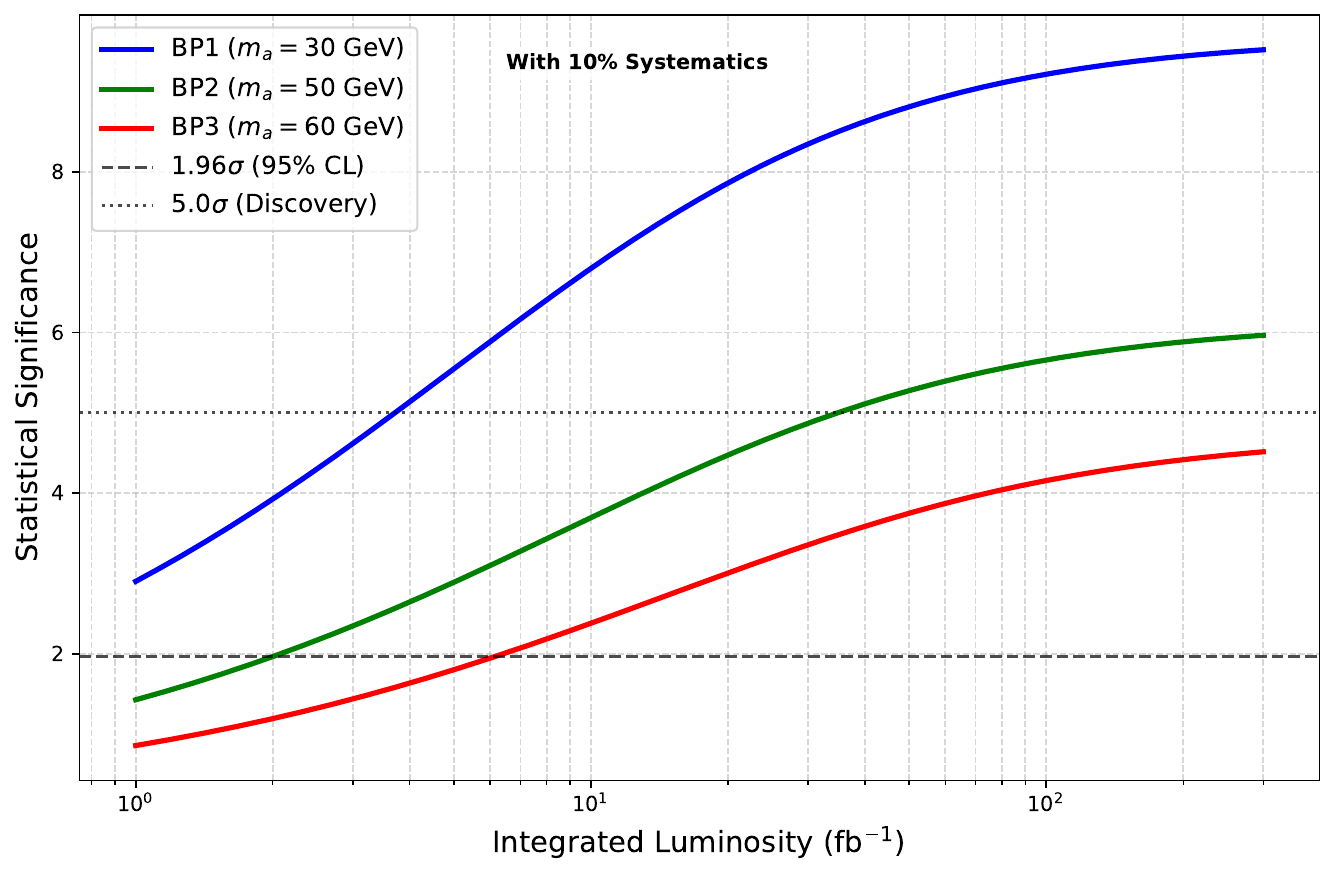}
  \put(0,22){\color{white}\rule{0.035\columnwidth}{0.29\columnwidth}}
  \put(1.5,36){\rotatebox{90}{\makebox(0,0){\sffamily\scriptsize Signal significance}}}
  \put(18.7,51.7){\color{white}\rule{0.095\columnwidth}{0.032\columnwidth}}
  \put(19,52.2){\sffamily\tiny (guide)}
 \end{overpic}
 \caption{Expected signal significance as a function of integrated luminosity for the three benchmark points at $\sqrt{s}=14$ TeV, including a 10\% systematic uncertainty in the background yield.}
 \label{fig:significance_lumi}
\end{figure}

\begin{table}[t]
 \caption{Expected signal significances for the three benchmark points at $\sqrt{s}=14$ TeV and the indicated integrated luminosities, including a 10\% systematic uncertainty in the background yield.}
 \label{tab:significance_vs_lumi}
 \centering
 \begin{ruledtabular}
 \begin{tabular}{rrrr}
 $\mathcal L\ (\fb)$ & BP1 & BP2 & BP3\\
 36 & 8.53 & 5.03 & 3.50\\
 100 & 9.21 & 5.66 & 4.15\\
 140 & 9.34 & 5.78 & 4.30
 \end{tabular}
 \end{ruledtabular}
\end{table}

Figure~\ref{fig:significance_lumi} and Table~\ref{tab:significance_vs_lumi} show the expected signal significance for the three benchmark masses with a 10\% systematic uncertainty in the background yield. For the adopted effective-vertex benchmark normalizations, the 30 and 50 GeV signals reach significances of 8.53 and 5.03, respectively, at $36~\fb$, exceeding the conventional $5\sigma$ discovery threshold. At $140~\fb$, their significances increase to 9.34 and 5.78. The 60 GeV benchmark reaches 3.50 at $36~\fb$ and 4.30 at $140~\fb$. The quoted significance levels, including the $5\sigma$ discovery threshold, are benchmark expectations for the reference effective-model normalization and do not imply an observed collider signal.

The intermediate luminosity of $100~\fb$ gives significances of 9.21, 5.66 and 4.15 for BP1, BP2 and BP3, respectively. Comparing these entries with those at $140~\fb$ shows that the gain over the last interval is smaller than over the preceding increase from $36~\fb$. The same behaviour appears in the gradual flattening of the curves in Fig.~\ref{fig:significance_lumi}. BP1 has the largest significance throughout the displayed range, followed by BP2 and BP3. This ordering accompanies the different reference signal yields and the selected signal-to-background ratios in Table~\ref{tab:yields}; it combines the assumed production rates with the acceptance and selection for each benchmark.

The luminosity dependence follows from the uncertainty prescription used in Eq.~\eqref{eq:significance}. Increasing $\mathcal L$ increases both selected event counts, while their ratio remains fixed for an unchanged selection. The absolute background uncertainty $\sigma_B=\epsilon B$ also increases with luminosity because $\epsilon$ is held at 0.1. Consequently, larger event samples progressively encounter a limitation from the assumed background normalization, and the significance rises more slowly. The curves therefore describe the joint effect of additional statistics and the fixed fractional uncertainty. A change in the background-uncertainty assumption would change the numerical sensitivity, even for the same tagged signal and background samples.

These results demonstrate the sensitivity of the squeezed double-$b$ strategy to the lightest benchmark and its applicability to the 50 and 60 GeV examples. The reduction of the background depends on combining displaced-track tagging, groomed-mass reconstruction and recoil kinematics. The tagger identifies the compact two-bottom candidate, the mass interval selects the resonance region, and the event BDT exploits correlations with the accompanying jets. Their combined response gives the selected rates used in the significance calculation, with the production normalization and background uncertainty stated explicitly.

\section{Summary and conclusions}
\label{sec:conclusion}

We have presented a tracking-based strategy for identifying squeezed double-$b$ jets from light hadronic resonances. When the two bottom-hadron decay systems lie within a single jet, their multiplicity is inferred from the distribution of displaced tracks. A three-category BDT combines displaced-track counts and momentum fractions with ordered impact parameters and jet kinematics to distinguish the $0b$, $1b$ and $2b$ configurations. This classification is the central reconstruction component of the study.

In the reference simulation, approximately $83\%$ of true double-bottom jets are assigned to the $2b$ category. The corresponding responses for true single-bottom and non-bottom jets are $9.7\%$ and $0.5\%$, respectively. The separate charm assessment shows that jets containing two charm hadrons can also contribute to the tagged double-bottom population. These category responses, together with the input jet populations, determine the composition after tagging and establish the importance of treating single-bottom and charm contamination explicitly.

The singlet-extended flipped 2HDM provides an illustrative signal model with pseudoscalar masses of 30, 50 and 60 GeV. The benchmarks are chosen primarily to examine the pseudoscalar mass dependence. The signal samples use a simplified effective-vertex approximation, in which the momentum-dependent quark-loop production interaction is represented by a local vertex. Soft-drop reconstruction preserves a localized mass structure near 30 GeV for the BP1 signal, while the tagged QCD background forms a continuum. The search combines this mass information with the double-$b$ tag and an event BDT using the candidate and recoil jets. At $\sqrt{s}=14$ TeV and $140~\fb$, the final signal-to-background ratios are approximately 1.30, 0.75 and 0.56 for the three masses, respectively. These ratios summarize the combined event selection, whose efficiency is distinct from the jet-category response.

For these benchmark signal rates, the expected significances at $140~\fb$ are 9.34, 5.78 and 4.30 for the three masses, respectively, assuming a $10\%$ background-normalization uncertainty. These are reference benchmark sensitivities; a resolved loop prediction and the corresponding higher-order production calculation are beyond the scope of the present tagging study. The reconstruction sequence can be applied to different signal hypotheses by evaluating their kinematic acceptance and tagging response and supplying the corresponding production rate. Squeezed double-$b$ tagging thus provides a common reconstruction approach for compact bottom-quark pairs in low-mass hadronic searches.

\begin{acknowledgments}
The authors acknowledge the use of the Kepler HPC facility at IISER Kolkata. S.S. thanks CSIR for funding. S.S. and R.K.S. acknowledge the hospitality of IACS, Kolkata, where part of the work was carried out. B.M. thanks the Department of Atomic Energy, Government of India, for support through a Raja Ramanna Chair position.
\end{acknowledgments}

\bibliographystyle{apsrev4-2}
\bibliography{refs}
\end{document}